\documentclass[traditabstract]{aa}

\usepackage{natbib}
\usepackage{graphicx}
\usepackage{txfonts}
\usepackage[bookmarks=false, colorlinks=true, citecolor=blue, linkcolor=blue]{hyperref}
\usepackage{subfig}

\def\micron{\hbox{\,$\mu$m}}
\newcommand{\Msun}{\hbox{$M_{\rm \odot}$}}
\newcommand{\degree}{\ensuremath{^\circ}}

\bibpunct{(}{)}{;}{a}{}{,}

\DeclareCaptionListOfFormat{subreformat}{#1#2}
\titlerunning{Excitation and acceleration of the ESO~320-G030 molecular outflow}
\authorrunning{Pereira-Santaella et al.}

\begin{document}

\title{Excitation and acceleration of molecular outflows in LIRGs: The extended ESO~320-G030 outflow on 200-pc scales}

\author{M.~Pereira-Santaella\inst{\ref{inst1}} \and L.~Colina\inst{\ref{inst1}} \and S.~Garc\'ia-Burillo\inst{\ref{inst2}} \and E.~Gonz\'alez-Alfonso\inst{\ref{inst3}} \and A.~Alonso-Herrero\inst{\ref{inst4}} \and S.~Arribas\inst{\ref{inst1}} \and S.~Cazzoli\inst{\ref{inst5}} \and J.~Piqueras-L\'opez\inst{\ref{inst1}} \and D.~Rigopoulou\inst{\ref{inst6}} \and A.~Usero\inst{\ref{inst2}}
}

\institute{Centro de Astrobiolog\'ia (CSIC-INTA), Ctra. de Ajalvir, Km 4, 28850, Torrej\'on de Ardoz, Madrid, Spain
\\ \email{miguel.pereira@cab.inta-csic.es}\label{inst1}
\and 
Observatorio Astron\'omico Nacional (OAN-IGN)-Observatorio de Madrid, Alfonso XII, 3, 28014, Madrid, Spain\label{inst2}
\and
Universidad de Alcal\'a, Departamento de F\'isica y Matem\'aticas, Campus Universitario, 28871 Alcal\'a de Henares, Madrid, Spain\label{inst3}
\and
Centro de Astrobiología (CSIC-INTA), ESAC Campus, E-28692 Villanueva de la Ca\~nada, Madrid, Spain\label{inst4}
\and
Instituto de Astrof\'isica de Andaluc\'ia, CSIC, Glorieta de la Astronom\'ia, s/n, E-18008 Granada, Spain\label{inst5}
\and
Department of Physics, University of Oxford, Keble Road, Oxford OX1 3RH, UK\label{inst6}
}

\abstract{
We used high-spatial resolution (70\,pc; 0\farcs3) CO multi-transition ($J$=1--0, 2--1, 4--3, and 6--5) ALMA data to study the physical conditions and kinematics of the cold molecular outflow in the local luminous infrared galaxy (LIRG) ESO~320-G030 ($d$=48\,Mpc, $L_{\rm IR}\slash L_\odot=10^{11.3}$). ESO~320-G030 is a double-barred isolated spiral, but its compact and obscured nuclear starburst (SFR$\sim$15\,\Msun\,yr$^{-1}$; $A_{\rm V}$\,$\sim$40\,mag) resembles those of ultra-luminous infrared galaxies ($L_{\rm IR}\slash L_\odot> 10^{12}$). In the outflow, the CO(1--0)\slash CO(2--1) ratio is enhanced with respect to the rest of the galaxy and the CO(4--3) transition is undetected. This indicates that the outflowing molecular gas is less excited than the molecular gas in the nuclear starburst (i.e., outflow launching site) and in the galaxy disk. 
Non-local thermodynamic equilibrium radiative transfer modeling reveals that the properties of the molecular clouds in the outflow differ from those of the nuclear and disk clouds: The kinetic temperature is lower ($T_{\rm kin}\sim$ 9\,K) in the outflow, and the outflowing clouds have lower column densities. Assuming a 10$^{-4}$ CO abundance, the large internal velocity gradients, 60$^{+250}_{-45}$\,km\,s$^{-1}$\,pc$^{-1}$, imply that the outflowing molecular clouds are not bound by self-gravity. All this suggests that the life-cycle (formation, collapse, dissipation) of the galaxy disk molecular clouds might differ from that of the outflowing molecular clouds which might not be able to form stars.
The low kinetic temperature of the molecular outflow remains constant at radial distances between 0.3 and 1.7\,kpc. This indicates that the heating by the hotter ionized outflow phase is not efficient and may favor the survival of the molecular gas phase in the outflow. 
The spatially resolved velocity structure of the outflow shows a 0.8\,km\,s$^{-1}$\,pc$^{-1}$ velocity gradient between 190\,pc and 560\,pc and then a constant maximum outflow velocity of about 700--800\,km\,s$^{-1}$ up to 1.7\,kpc. This could be compatible with a pure gravitational evolution of the outflow, which would require coupled variations of the mass outflow rate and the outflow launching velocity distribution. Alternatively, a combination of ram pressure acceleration and cloud evaporation could explain the observed kinematics and the total size of the cold molecular phase of the outflow.

}
\keywords{galaxies: ISM -- galaxies: nuclei -- infrared: galaxies -- ISM: molecules}

\maketitle
 
\section{Introduction}\label{sec:intro}

\begin{table*}[t]
\caption{Summary of the ALMA observations}
\label{tbl:alma_obs}
\centering
\begin{small}
\begin{tabular}{ccccccccc}
\hline \hline
\\
Band &  Beam & Channel & Sensitivity\tablefootmark{a} & CO transition & Project  \\
& FWHM & (km\,s$^{-1}$) & (mJy\,beam$^{-1}$) & \\
\hline
\\
3 & 0\farcs37$\times$0\farcs30 & 5.1 & 0.87 & 1--0 & 2016.1.00263.S\\
6 & 0\farcs30$\times$0\farcs24 & 5.1 & 0.98 & 2--1 & 2013.1.00271.S\\
8 & 0\farcs28$\times$0\farcs27 & 5.1 & 4.2  & 4--3 & 2016.1.00263.S\\
9 & 0\farcs20$\times$0\farcs18 & 21  & 7.6  & 6--5 & 2016.1.00263.S\\
\hline
\end{tabular}
\end{small}
\tablefoot{
\tablefoottext{a}{1$\sigma$ sensitivity per spectral channel.}}
\end{table*}

Galaxy-scale massive gas outflows play a key role in the regulation and suppression of star-formation in galaxies according to numerical galaxy-formation simulations. These gas outflows are the result of energetic processes (e.g., strong radiation pressure, radio jets, supernovae, stellar winds) associated with active galactic nuclei (AGN) and starbursts (e.g., \citealt{Hopkins2014, Schaye2015, Dubois2016, Dave2019, Nelson2019, Mitchell2020}).

Observationally, outflows are ubiquitous in active galaxies (e.g., \citealt{Fiore2017, Rupke2018, Veilleux2020}) and present a multiphase structure with atomic ionized and neutral gas (e.g., \citealt{Perna2017, RodriguezdelPino2019}) as well as cold, warm, and hot molecular gas (e.g., \citealt{Spoon2013, Hill2014, Emonts2017, Fluetsch2019}). However, the cold molecular phase is typically the most massive (e.g., \citealt{Fiore2017, Spence2018}).

In this paper, we focus on the cold ($T < 100$\,K) molecular phase of the outflows, which can be traced by the low-$J$ CO transitions. CO outflows have been observed in numerous local AGN (e.g., \citealt{Combes2013, GarciaBurillo2014, Dasyra2014, Morganti2015, AAH2018, AAH2019, FernandezOntiveros2020}) and starbursts (e.g., \citealt{Walter2002, Bolatto2013ngc253, Pereira2016b, Krieger2019}). In addition, high-velocity CO outflows are detected in local ultra-luminous infrared (IR) galaxies (ULIRGs; $L_{\rm IR}\slash L_\odot > 10^{12}$), where compact and obscured starbursts, sometimes together with an AGN, power strong molecular outflows with mass outflow rates comparable to the star-formation rate of the host object (e.g., \citealt{Feruglio2010, Cicone2014, Pereira2018, BarcosMunoz2018, Lutz2020}).
However, most of these studies only observe one CO transition, and the molecular outflow is not always fully spatially resolved.
Therefore, some properties of the molecular gas evolution within the outflow (e.g., destruction and formation timescales, acceleration, thermal evolution), which are important to interpreting local and high-$z$ unresolved outflow observations, are not well constrained. To investigate this, we obtained high-spatial resolution (70\,pc; 0\farcs3) CO multi-transition (1--0, 2--1, 4--3, and 6--5) ALMA observations of the local luminous IR galaxy (LIRG) ESO~320-G030 ($d=48$\,Mpc; $L_{\rm IR}\slash L_\odot = 10^{11.3}$).

ESO~320-G030 (or IRAS~F11506-3851) is an isolated double-barred spiral with an ordered gas velocity field \citep{Bellocchi2013, Bellocchi2016, Pereira2016b} known to host a fast and massive multiphase outflow; its atomic neutral \citep{Cazzoli2014}, atomic ionized \citep{Arribas2014}, and molecular \citep{Pereira2016b} phases have already been studied. 
This outflow originates at the nucleus and its orientation is approximately perpendicular to the stellar disk. The nuclear region (central $\sim$250\,pc)  hosts a strong obscured starburst (nuclear SFR$\sim$15\,\Msun\,yr$^{-1}$; $A_{\rm V}$\,$\sim$40\,mag), which produces half of the total IR luminosity. There is no evidence for a bright AGN based on mid-IR and X-ray observations \citep{Pereira2010, Pereira2011, AAH2012a}, although the modeling of the far-IR molecular absorptions is compatible with a low-luminosity ($<10\%$ of the total $L_{\rm IR}$) and heavily embedded ($N_{\rm H}>7\times 10^{24}$\,cm$^{-2}$) AGN (Gonz\'alez-Alfonso et al. 2020 submitted). The energy and momentum of the observed outflow can be explained by the  compact nuclear starburst (see \citealt{Pereira2016b}).
The nuclear molecular gas surface density is very high: $\Sigma_{\rm H_2} = 10^{4.4}$\,$M_\odot$\,pc$^{-2}$. It is comparable to the $\Sigma_{\rm H_2}$ observed in more extreme ULIRGs (e.g., \citealt{Wilson2019}), about 200 times higher than the values observed in nearby disk galaxies (e.g., \citealt{Bigiel2011}), and $\sim$10 times higher than in local Seyfert galaxies (e.g., \citealt{AAH2020}).
Therefore, this object allows us to carry out a detailed study of the properties of outflows produced in compact and obscured nuclear regions similar to those of more distant starburst-dominated ULIRGs.

We organize this paper as follows. In Section~\ref{s:data}, we present the ALMA observations and data reduction. Section~\ref{s:data_analysis} describes how the spatially resolved CO spectral line energy distributions (SLEDs) are obtained and how they are modeled using non-local thermodynamic equilibrium (non-LTE) radiative transfer models. In Section~\ref{s:evol}, we discuss the thermal and dynamic evolution of the outflow cold molecular phase. Finally, in Section~\ref{s:conclusions}, the main findings of this paper are summarized.

\section{ALMA data reduction}\label{s:data}

We observed the 1--0, 2--1, 4--3, and 6--5 CO transitions with the ALMA 12-m array (Bands 3, 6, 8, and 9, respectively) in the galaxy ESO~320-G030. These observations were obtained through 2 programs: Bands 3, 8, and 9 from 2016.1.00263.S (PI: M. Pereira-Santaella) and Band 6 from 2013.1.00271.S (PI: L. Colina). The Band 6 CO(2--1) data have been already analyzed by \citet{Pereira2016b}, and the first detection in space of the H$_2$O 448\,GHz transition (Band 8 data) was presented in \citet{Pereira2017Water}.

Two 12-m array configurations, compact and extended, were considered to obtain the same angular resolution ($\sim$0\farcs30;  70\,pc) and maximum recoverable scale (3\arcsec; 700\,pc) for all the CO transitions, except for the higher frequency Band 9 data for which only the extended configuration was used. The Band 9 ALMA Compact Array observation was scheduled but it did not pass the quality assurance and was not combined with the 12-m array data.
The Band 3 and 6 observations were single pointing centered at the nucleus of ESO~320-G030. To cover the central $\sim$10\arcsec\ of the target for all the CO transitions, the Band 8 and 9 observations required three and five pointing mosaics, respectively.

Four spectral windows of 1.875\,GHz were defined for all the Bands centered on various molecular transitions and continuum. In this paper, we focus on the spectral windows centered on the CO transitions: CO(1--0) 115.271\,GHz, CO(2--1) 230.538\,GHz, CO(4--3) 461.041\,GHz, and CO(6--5) 691.473\,GHz, which have spectral channels of $\sim$1--3\,km\,s$^{-1}$.

We calibrated the data using \textsc{CASA} (v5.4). For the bands with two array configurations (all except Band 9), we verified that the amplitudes for the baselines in common for both configurations were in good agreement.
The continuum was subtracted in the {\it uv} plane by fitting a constant value using the line-free channels.
The data were then cleaned using the \textsc{tclean} \textsc{CASA} task and the Brigss weighting \citep{Briggs1995PhDT} with a robustness parameter of 0.5. The final cleaned cubes have 5\,km\,s$^{-1}$ channels (except the Band 9 that has 21\,km\,s$^{-1}$ channels) and 0\farcs06 (0\farcs04 for Band 9) pixels, which properly sample the beam. These cubes were corrected for the primary beam patterns. The recovered beams and sensitivities are indicated in Table\,\ref{tbl:alma_obs}.

We assumed the nominal ALMA absolute flux accuracy in Cycle 4 for our data. That is, 5\% for Band 3, CO(1--0), 10\% for Band 6, CO(2--1), and 20\% for Bands 8 and 9, CO(4--3) and CO(6--5).

\subsection{HNCO emission subtraction}\label{ss:hnco}

\begin{figure}
\centering
\vspace{5mm}
\includegraphics[width=0.47\textwidth]{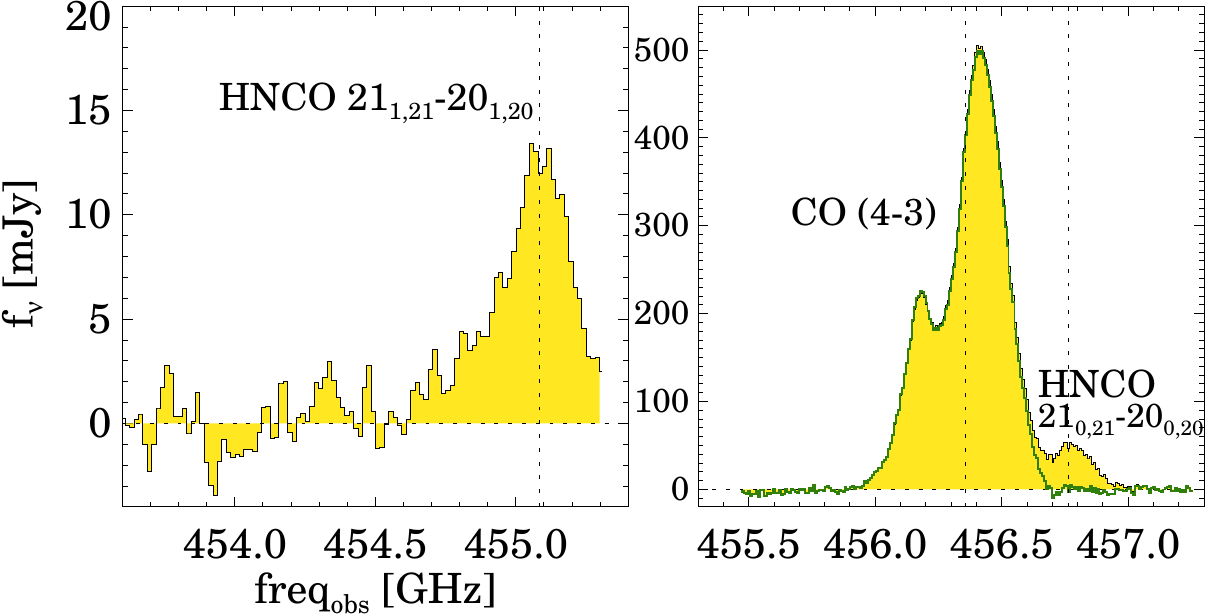}
\caption{ALMA Band 8 nuclear spectrum of ESO~320-G030 ($r=$0\farcs3). Each panel shows one of the spectral windows where two HNCO transitions and CO(4--3) are detected (see Section~\ref{ss:hnco}). We note that part of the blue wing of the HNCO $21_{1,21}-20_{1, 20}$ transition (left panel) is not covered by the observed spectral window.
The vertical dashed lines mark the expected frequency of the HNCO and CO transitions based on the systemic velocity. The green histogram in the right panel shows the nuclear CO(4--3) spectrum after subtracting the HNCO emission.\label{fig:hnco}}
\end{figure}

In the Band 8 spectrum of ESO~320-G030, we detected an emission line coincident with the CO(4--3) blue wing centered at $-270$\,km\,s$^{-1}$ with respect to the CO(4--3) systemic velocity (see Figure~\ref{fig:hnco} right panel). This line is only detected in the nuclear region ($r<$0\farcs3). Since we are interested in the high-velocity outflow emission in the CO line wings, we subtracted the emission of this transition.
We identified this transition as HNCO $21_{0,21}-20_{0, 20}$ 461.450\,GHz based on its observed frequency and also on the detection of the HNCO $21_{1,21}-20_{1, 20}$ 459.755\,GHz transition in another Band 8 spectral window (Figure~\ref{fig:hnco} left panel). Both HNCO transitions show a rotation pattern compatible with that observed in CO. To subtract the HNCO $21_{0,21}-20_{0, 20}$ emission, we first created a model of the HNCO $21_{1,21}-20_{1, 20}$ emission by fitting a 2D Gaussian to the image of each spectral channel. Then, we scaled this model to match the flux of the HNCO $21_{0,21}-20_{0, 20}$ transition. Finally, we subtracted the scaled HNCO model to the CO(4--3) cube. By doing this, most of the HNCO emission is removed leaving only small residuals (see Figure~\ref{fig:hnco} right). However, we note that the high-velocity blue-shifted nuclear CO(4--3) emission might be affected by these HNCO residuals, so we increased the uncertainties of these channels by the standard deviation of the residuals.

\subsection{Beam matching}\label{ss:beam_match}

The CO(1--0) data have a slightly worse beam than the other transitions (see Table~\ref{tbl:alma_obs}). Therefore, we degraded the angular resolution of the other CO data cubes to match the resolution of the CO(1--0). To do so, we used the \textsc{imsmooth} \textsc{CASA} task to obtain cubes with a 0\farcs37$\times$0\farcs30 (85\,pc$\times$70\,pc) beam.
In the following sections, we analyze these beam matched CO data cubes.

\section{Data analysis}\label{s:data_analysis}

\begin{figure*}[!htp]
\centering
\vspace{5mm}
\includegraphics[width=\textwidth]{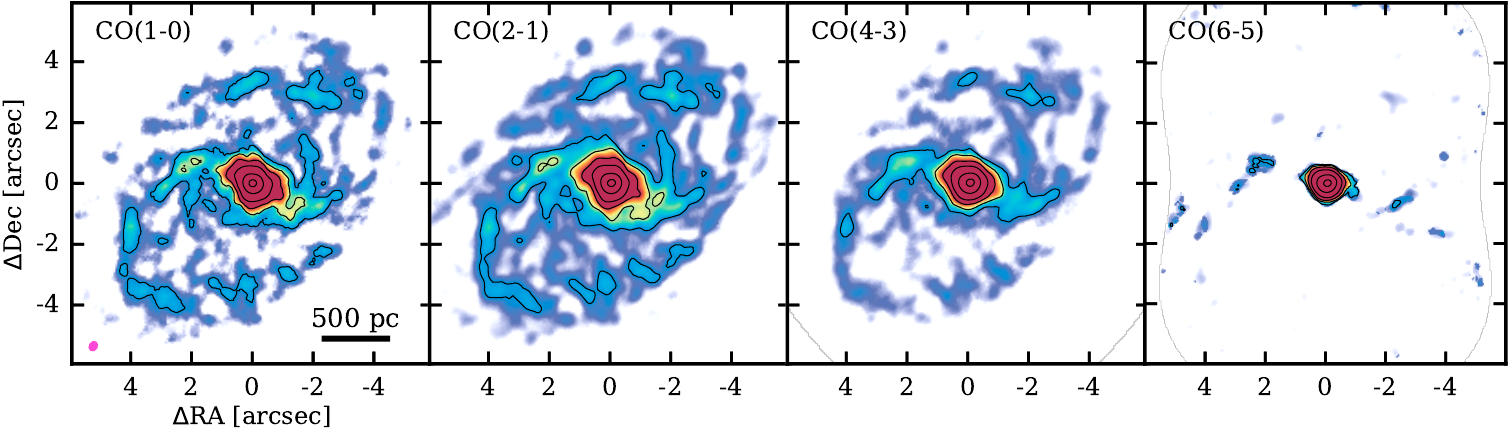}
\caption{CO 1--0, 2--1, 4--3, and 6--5 line emission maps (zero moment maps) of ESO~320-G030. The emission is dominated by the nuclear regions, so we saturated the color scale at 10\% of the emission peak to reveal the secondary bar and the internal spiral arms. The black contour levels are 0.02, 0.05, 0.1, 0.2, 0.5, 0.9 times the emission peak in each map (13.8, 61, 182, 242~Jy\,km\,s$^{-1}$\,beam$^{-1}$ for the 1--0, 2--1, 4--3, and 6--5 panels, respectively). The pink ellipse in the first panel represents the beam size (0\farcs37$\times$0\farcs30; PA $-32$\degree), which is common for the 4 panels (see Section~\ref{ss:beam_match}). The physical scale at the distance of this galaxy is indicated in the first panel. The gray lines in the CO(4--3) and CO(6--5) panels indicate the shape of the primary beam. For the CO(1--0) and CO(2--1) panels, the primary beam is larger than the field of view presented in this figure.
\label{fig:moment0}}
\end{figure*}

\begin{figure*}
\centering
\vspace{5mm}
\includegraphics[width=0.9\textwidth]{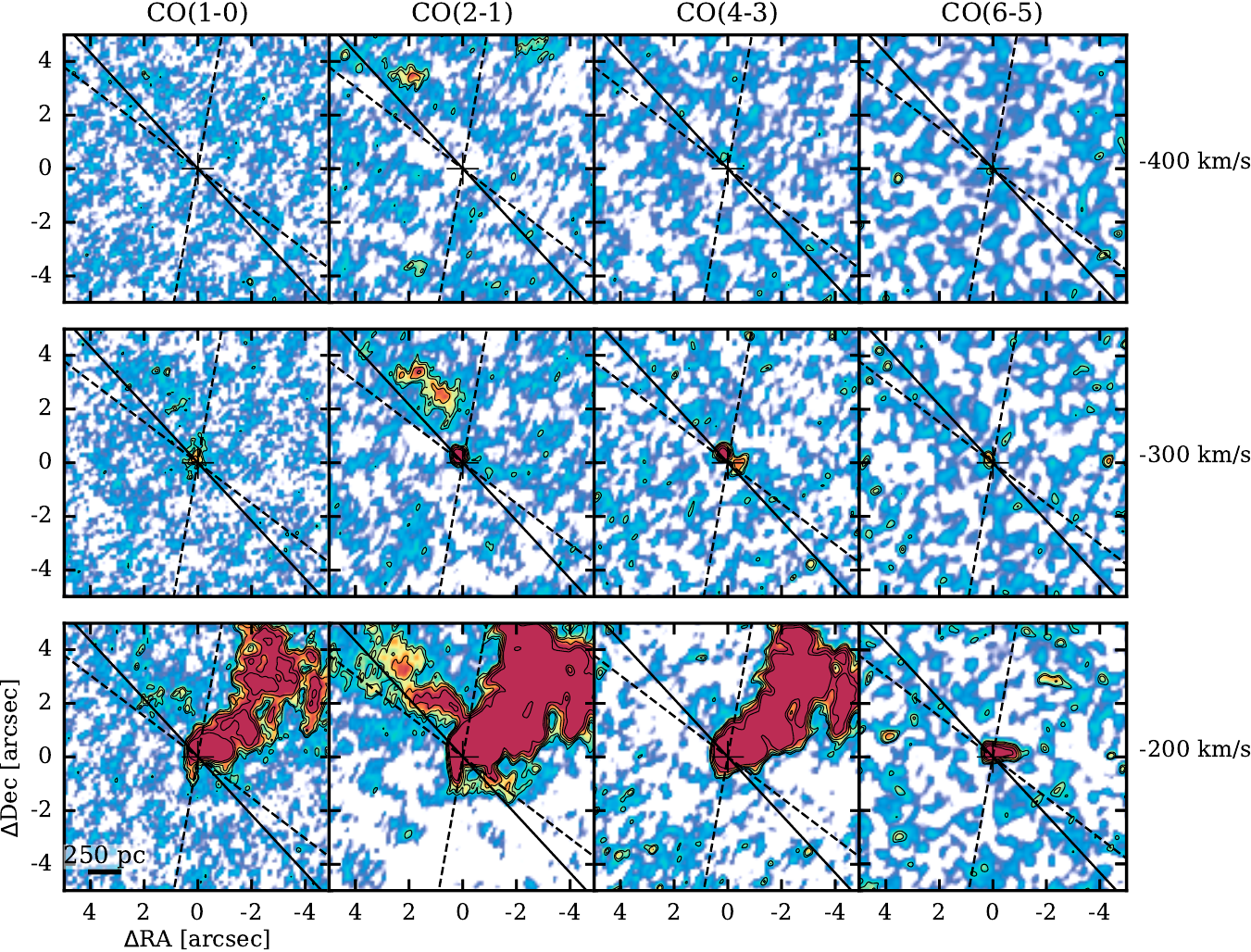}
\caption{Channel maps of the CO 1--0, 2--1, 4--3, and 6--5 transitions (from left to right) centered at observed velocities blue-shifted with respect to systemic by $-400$, $-300$, and $-200$\,km\,s$^{-1}$ (from top to bottom). The channel width is 100\,km\,s$^{-1}$. The contour levels are 3, 4, 6, 9, 13 times the $\sigma$ of each channel map (0.2, 0.2, 0.9, 4.4\,mJy\,beam$^{-1}$). The color scale is saturated at 9$\sigma$ to highlight emission at high velocities associated with the outflow.
The solid black line is kinematic minor axis (PA$=$43\degree) derived from the kinematic model \citep{Pereira2016b}. The dashed lines delimit the area where the high-velocity molecular outflow is detected (PAs from -10\degree to 53\degree). The black cross marks the position of the nucleus derived from the ALMA continuum emission (see \citealt{Pereira2017Water}). The physical scale at the distance of this galaxy is indicated in the lower left panel. \label{fig:channels_blue}}
\end{figure*}

\begin{figure*}
\centering
\vspace{5mm}
\includegraphics[width=0.9\textwidth]{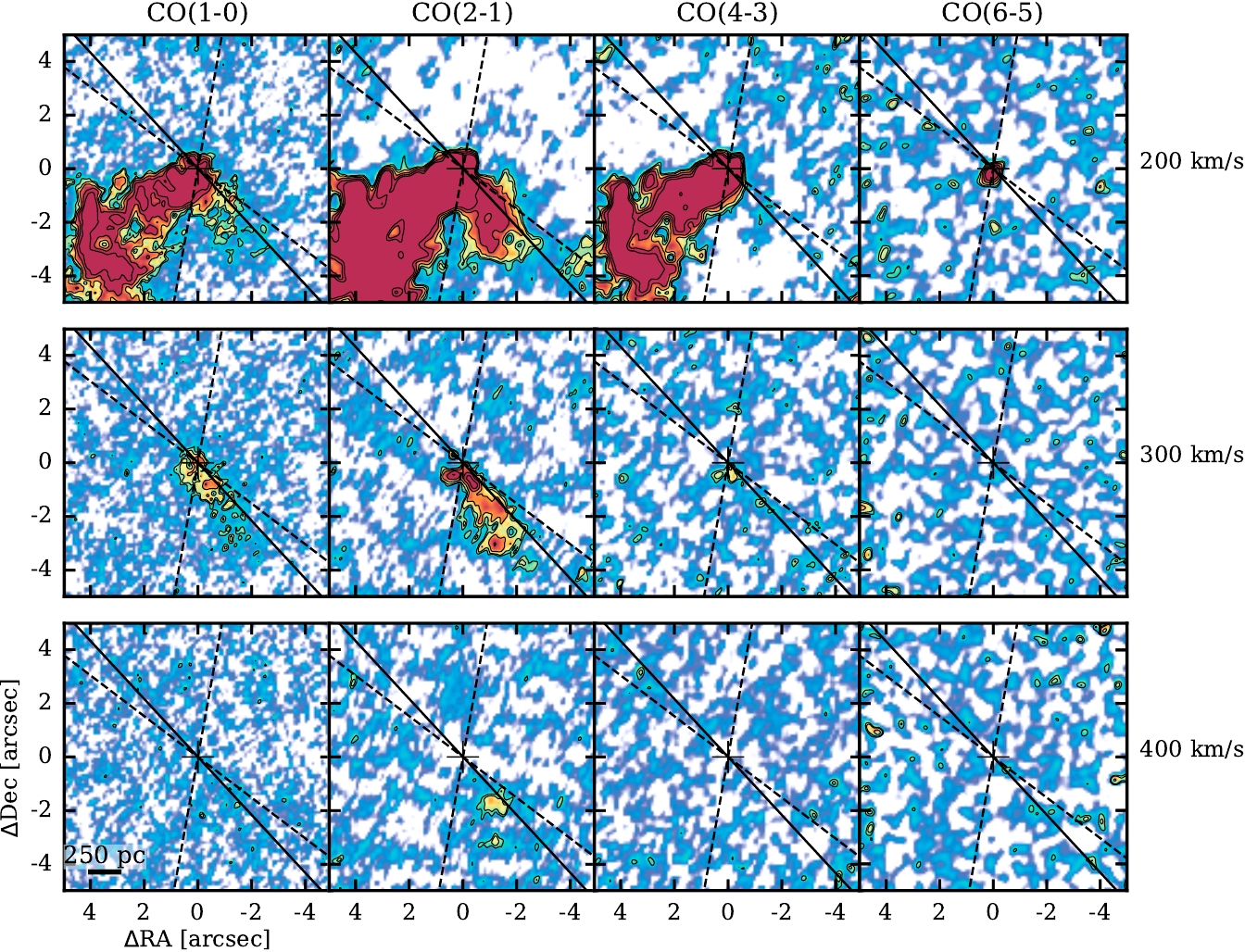}
\caption{Same as Figure~\ref{fig:channels_blue} but for the red-shifted emission, $+$200, $+$300, and $+$400\,km\,s$^{-1}$ (from top to bottom) with respect to systemic velocity. \label{fig:channels_red}}
\end{figure*}

\subsection{High-velocity gas morphology}

A spatially resolved (deprojected $r\sim$1.3\,kpc) high-velocity (up to 750\,km\,s$^{-1}$ corrected for inclination) molecular outflow has been detected in ESO~320-G030 (see \citealt{Pereira2016b}). The cold molecular phase is traced by the CO(2--1) 230\,GHz transition and the hot molecular gas by the near-IR H$_2$ 1--0 S(1) 2.12\micron\ transition. This outflow is approximately perpendicular to the galaxy disk and it originates within the central $\sim$250\,pc due to an intense (SFR$\sim$15\,\Msun\,yr$^{-1}$) and heavily obscured 
nuclear starburst.

In this section, we present new ALMA observations of the cold molecular gas traced by the CO transitions 1--0, 4--3, and 6--5. We combine the new data with the CO(2--1) observations already published by \citet{Pereira2016b}. In Figure~\ref{fig:moment0}, we present the integrated flux maps for the four CO transitions. These maps show the secondary nuclear bar and the internal spiral structure. These nuclear bar and spiral arms are located within the large scale primary bar (9~\,kpc semi-major axis; see \citealt{Pereira2016b} for more details).

Figures~\ref{fig:channels_blue} and \ref{fig:channels_red} show the channel maps of the blue- and red-shifted emissions (observed $\pm$400, $\pm$300, and $\pm$200\,km\,s$^{-1}$) with respect to the systemic velocity (v$_{\rm radio} = $3049\,km\,s$^{-1}$; \citealt{Pereira2017Water}). In the lowest velocity maps ($\pm$200\,km\,s$^{-1}$), the disk emission is clearly detected in the lower-$J$ CO transitions (1--0, 2--1, and 4--3), but not in CO(6--5). At these velocities, the CO(6--5) is only detected in the central $\sim$250\,pc.

The outflow emission is detected in the higher velocity panels ($|$v$|>200$\,km\,s$^{-1}$) approximately along the kinematic minor axis (solid line). This orientation of the high-velocity gas is compatible with an outflow perpendicular to the galaxy disk.
The extended ($r$ up to 5\arcsec) high-velocity outflow emission is detected through the CO(2--1) and CO(1--0) transitions. In contrast, no large-scale high-velocity CO(4--3) and CO(6--5) emissions are detected.

\subsection{Spatially resolved outflow CO SLED}\label{ss:outflow_sled}

\begin{figure*}[!htp]
\centering
\vspace{5mm}
\includegraphics[width=0.9\textwidth]{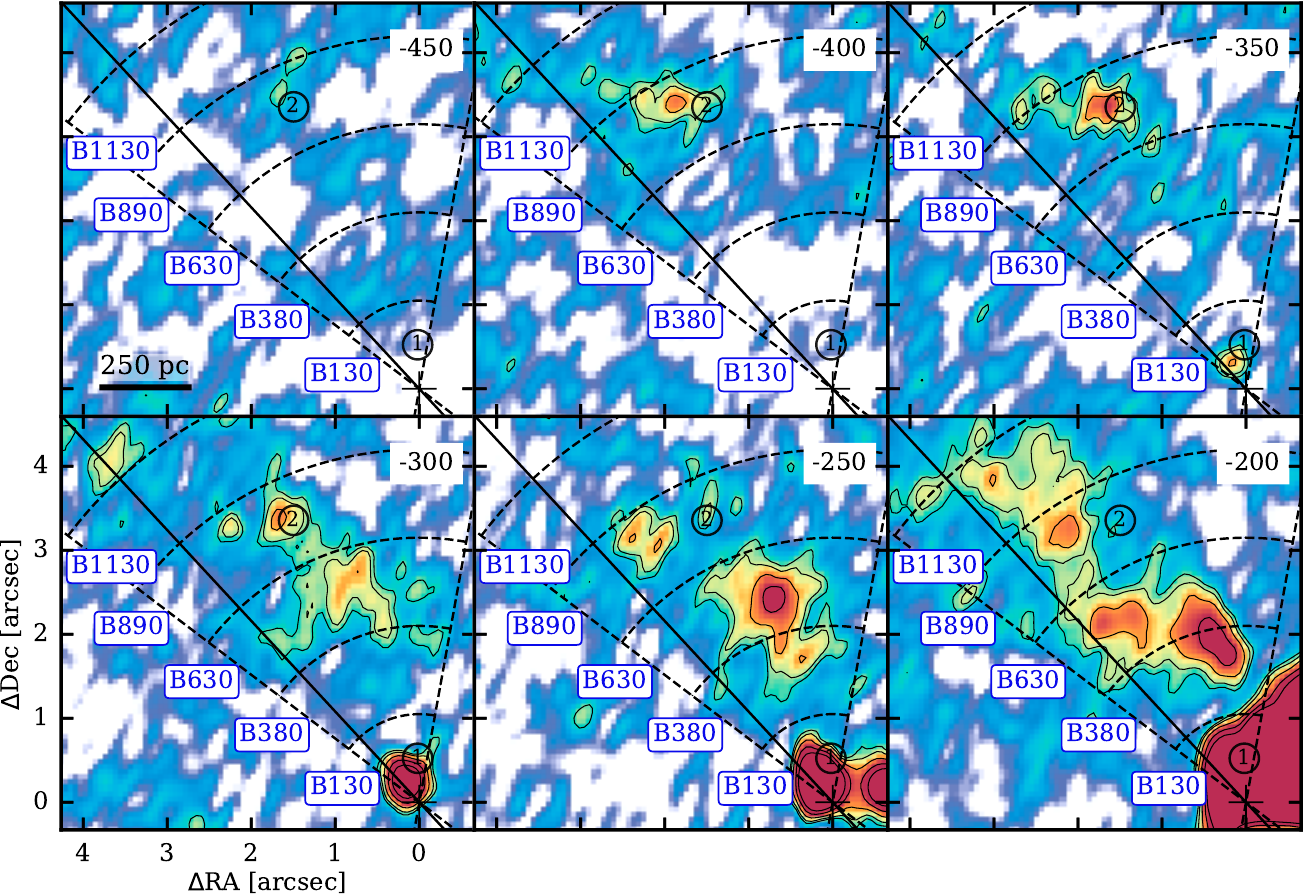}
\includegraphics[width=0.85\textwidth]{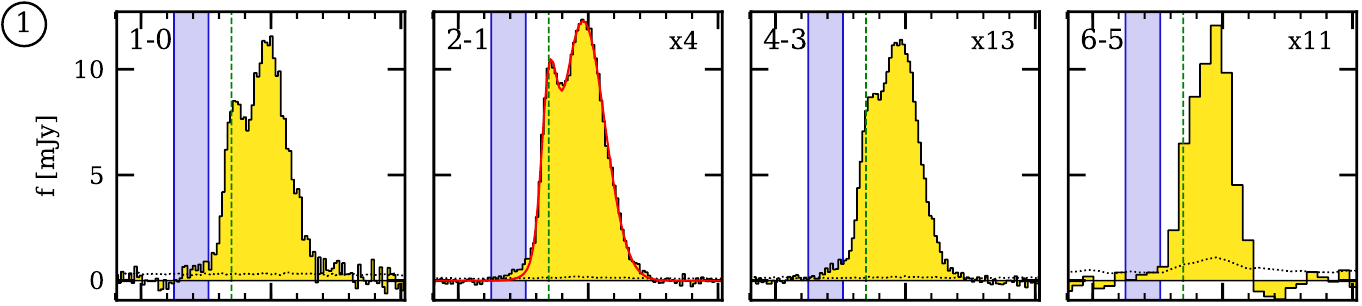}
\includegraphics[width=0.85\textwidth]{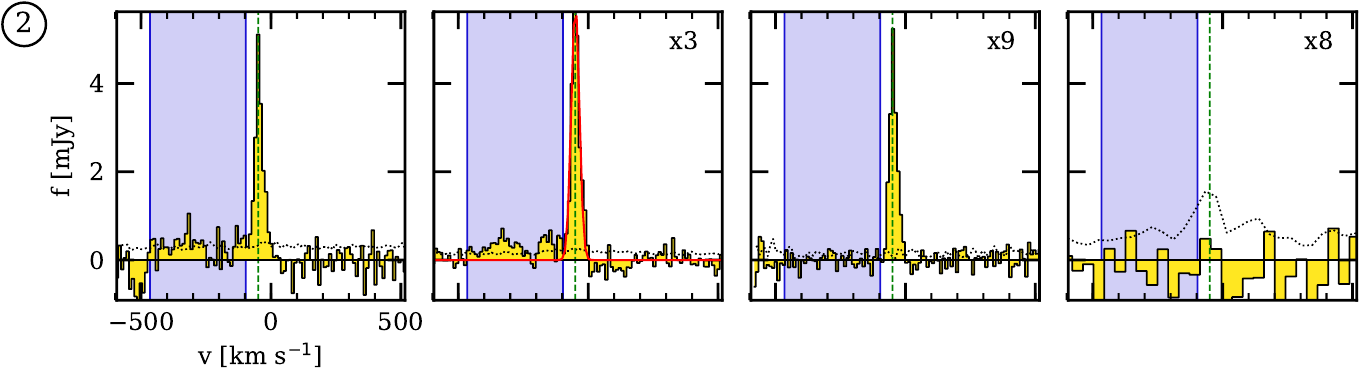}
\caption{Top panels: high-velocity blue-shifted CO(2--1) channel maps (v$_{\rm los}$ from  $-$450 to $-$200\,km\,s$^{-1}$ as indicated at the top right corner of each map). The black solid line is the kinematic minor axis of the galaxy and the radial dashed lines mark the angular extent of the outflow emission (see Figure~\ref{fig:channels_blue}). The dashed arcs delimit the annular sectors used to measure the resolved outflow CO SLEDs. The B130 to B1130 labels indicate the projected separation in parsec between the nucleus and the centers of the annulus sectors.
Bottom panels: CO(1--0), CO(2--1), CO(4--3), and CO(6--5) integrated spectra of two regions as an example (see Section~\ref{ss:outflow_sled}). The location of these two regions is marked in the top maps with a circled number in the B130 and B890 sectors. The spectra are scaled by the factors at the upper right corner of each panel.
The dashed green vertical lines mark the disk rotation velocity derived from the kinematic model at each position (Section~\ref{ss:outflow_sled}). The red line in the CO(2--1) panel is the best fitting one- or two-Gaussian model. The blue shaded area indicates the spectral channels that are used to measure the high-velocity CO molecular emission. The dotted black line is the 1$\sigma$ noise level measured in each channel.
\label{fig:annuli_blue}}
\end{figure*}

\begin{figure*}[!htp]
\centering
\vspace{5mm}
\includegraphics[width=0.9\textwidth]{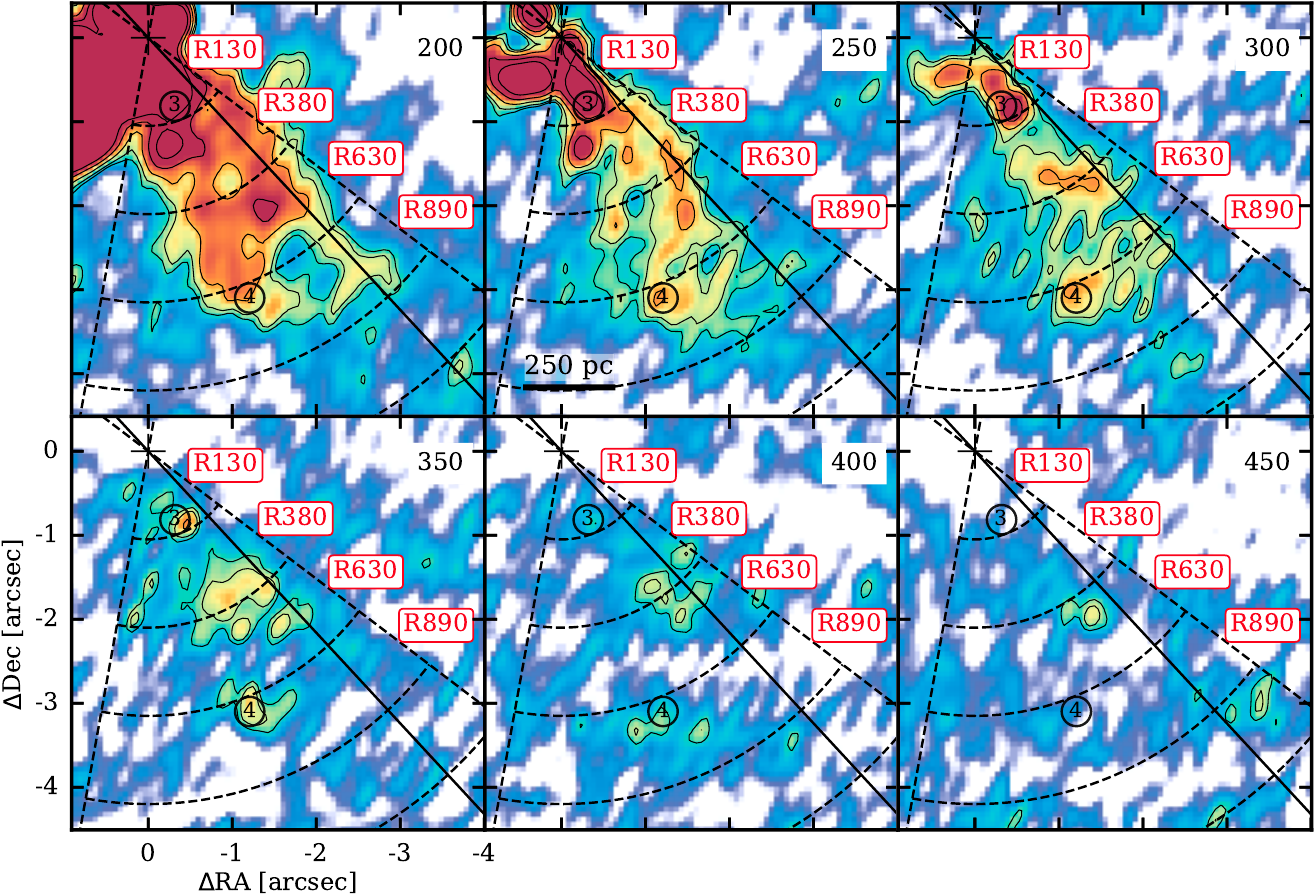}
\includegraphics[width=0.85\textwidth]{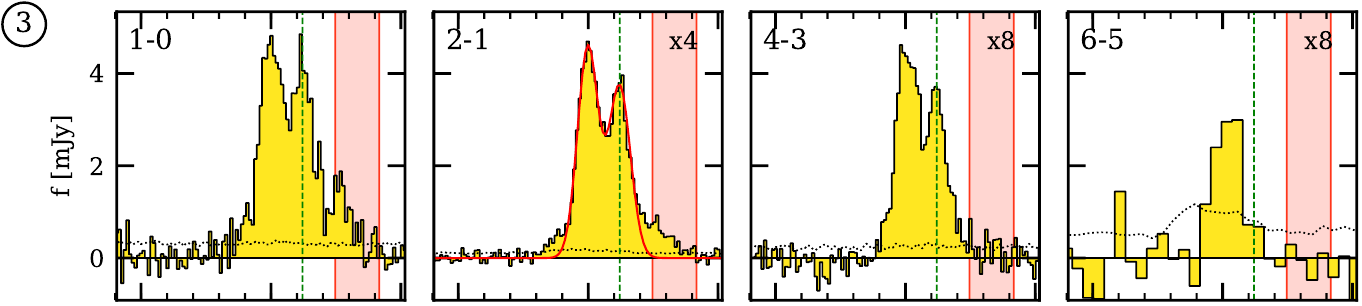}
\includegraphics[width=0.85\textwidth]{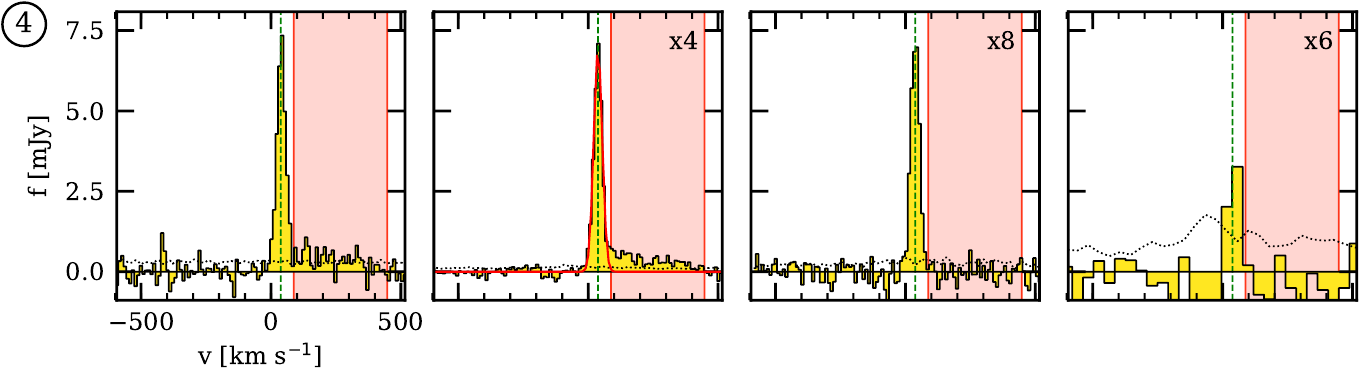}
\caption{Same as Figure~\ref{fig:annuli_blue} but for the red-shifted side of the outflow. The location of the two regions whose spectra are shown in the bottom panels is marked in the top maps with a circled number in the R130 and R890 sectors. \label{fig:annuli_red}}
\end{figure*}

To obtain the spatially resolved CO spectral line energy distribution (SLED) of the outflow, we first defined the outflow region as a biconical structure projected on the plane of the sky (marked by dashed lines in Figures~\ref{fig:channels_blue} and \ref{fig:channels_red}) which encompasses the emission of the high-velocity molecular gas. The opening angle of the outflow region is 65\degree\ and its central PA is 22\degree\ (east of north).
This PA differs by $\sim$20\degree\ from the kinematic minor axis and suggests some deviation from an orientation perfectly perpendicular to the disk. 
In this section, all the velocities and distances referred to the outflow are line of sight velocities and distances projected on the sky. The inclination of ESO~320-G030 is 43$\pm$2\degree\ relative to the plane of the sky \citep{Pereira2016b}, so, assuming an orientation perpendicular to the disk, the velocities and distances should be multiplied by $1\slash \cos i\sim 1.4$ and $1\slash \sin i\sim 1.5$, respectively, to obtain the actual outflow values.

Then, we constructed contiguous annulus sectors with a width of $\sim$1\farcs1, which corresponds to 3 times the beam FWHM ($\sim$250\,pc). The innermost sector starts at the galaxy nucleus and the most external one reaches the more distant position where high-velocity gas is detected (1.3 and 1.1\,kpc for the blue- and red-shifted emission, respectively). In total, we defined 6 annulus sectors for the blue-shifted outflow emission and 5 sectors for the red-shifted outflow emission (see top panels of Figures~\ref{fig:annuli_blue} and \ref{fig:annuli_red}).

Within these annulus sectors, we measured the CO emission in nonoverlapping circular regions with a diameter of 0\farcs35 (i.e., approximately the beam FWHM). The spectra of four characteristic regions are shown in the bottom panels of Figures~\ref{fig:annuli_blue} and \ref{fig:annuli_red}.

To identify the high-velocity gas, we first determined the line-of-sight velocity of the galaxy disk molecular gas at each position. To do so, we used the CO(2--1) kinematic model described in \citet{Pereira2016b}. This model consists of a thin rotating disk with the azimuthal velocity varying as a function of the radius (i.e., rotation curve) and no velocity in the radial or perpendicular to the disk directions. The rotation curve (Figure 5 of \citealt{Pereira2016b}) is obtained by fitting the observed CO(2--1) velocity field.

\citet{Pereira2016b} derived the CO(2--1) velocity field in the central 12\arcsec ($\sim$3\,kpc) by fitting a single Gaussian model to the CO emission profiles. However, they excluded the nuclear CO emission (r$<$1\arcsec; $<$230\,pc) since it shows a more complex profile that is not well reproduced by a single Gaussian. 
To measure the high-velocity gas in the nuclear region, we need to expand the velocity field to include the nucleus too. Therefore, we used a two-Gaussian model to reproduce the nuclear profiles (see spectra of regions 1 and 3 in Figures~\ref{fig:annuli_blue} and \ref{fig:annuli_red}). With this two-Gaussian model, we were able to account for most of the emission in the central region of ESO~320-G030. 
One component corresponds to molecular gas with inward radial velocities associated with the nuclear secondary bar; and the second component to a nuclear rotating molecular disk. A more detailed physical interpretation of these two components is presented in another work (Gonz\'alez-Alfonso et al. 2020 submitted).  

From the kinematic model, we obtained the velocity, ${\rm v}_{\rm disk}$, and velocity dispersion, $\sigma_{\rm disk}$, of the molecular gas in the disk of the galaxy at each position.
To determine the velocity range of the high-velocity CO(2--1) emission, we used the ${\rm v}_{\rm disk}$ and $\sigma_{\rm disk}$ in each aperture.
This high-velocity range starts at ${\rm v}_{\rm disk} + 3 \sigma_{\rm disk}$, and at ${\rm v}_{\rm disk} - 3 \sigma_{\rm disk}$ for the red- and blue-shifted sides, respectively.
Then, we extended the high-velocity range until the flux of the channels, when convolved to a resolution of half the width of the range, was below the detection limit (see the blue- and red-shaded areas in the bottom panels of Figures~\ref{fig:annuli_blue} and \ref{fig:annuli_red}).
We measured the high-velocity CO emission for all the transitions using the velocity range defined from the CO(2--1) profile, which has the higher signal-to-noise ratio.

This method works well for regions far from the nucleus (regions 2 and 4 in Figures~\ref{fig:annuli_blue} and \ref{fig:annuli_red}, respectively) but, close to the nucleus, the high-velocity emission is likely underestimated and should be considered as a lower limit. This is because the nuclear CO line profile is very broad (FWHM$\sim$250\,km\,s$^{-1}$) and the emission from outflowing gas at relatively low velocities ($\sim$200\,km\,s$^{-1}$), which is clearly detected at larger radii is blended with the nuclear profile. However, we are interested in the radial variations of the CO SLED. Assuming that the SLED does not depend on the velocity of the gas within a given sector, we can use the fluxes measured in the inner sector to derive the CO SLED even if they do not account for the lowest velocity outflowing gas.

\begin{figure*}[t]
\centering
\vspace{5mm}
\includegraphics[width=0.95\textwidth]{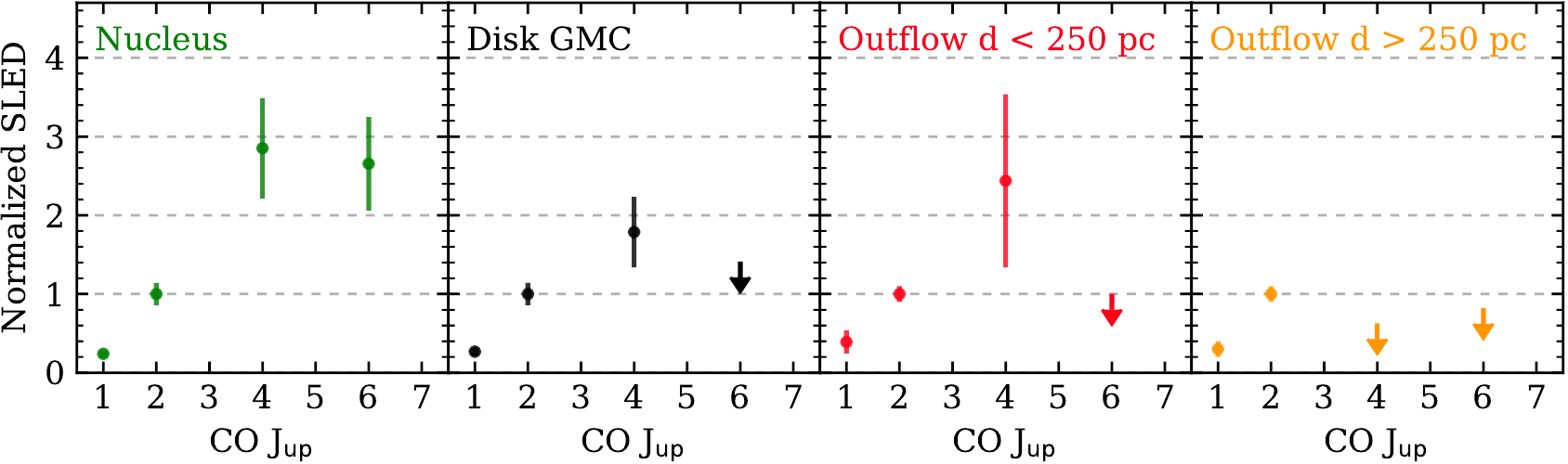}
\caption{CO SLEDs normalized to the flux of the CO(2--1) transition. The nuclear SLED corresponds to the central 100\,pc (diameter$=$0\farcs4). The Disk GMC panel shows the average CO SLED of the gas in the disk and the two outflow panels show the average SLEDs for the inner annulus sector (distance $<250$\,pc) and the rest of sectors, respectively (see Figures~\ref{fig:annuli_blue} and \ref{fig:annuli_red}).
The uncertainties include both flux uncertainties and the spatial variations of the SLED within each region.\label{fig:co_sled}}
\end{figure*}

\begin{table*}[!h]
\caption{CO fluxes}
\label{tbl:co_fluxes}
\centering
\begin{small}
\begin{tabular}{lcccccccc}
\hline \hline
\\
Region & Area\tablefootmark{a} & CO(1--0) & CO(2--1) & CO(4--3) & CO(6--5) \\
&(arcsec$^2$) & \multicolumn{4}{c}{(Jy km\,s$^{-1}$)} \\
\hline
Nucleus\tablefootmark{b}  & 4.4 &   81.7 $\pm$ 0.4  &  340.5 $\pm$ 1.9 & 970 $\pm$ 190  &  900 $\pm$ 180  \\
Disk GMCs  & 39 &   78.7 $\pm$ 0.4  &  307.3 $\pm$ 1.5 &  538 $\pm$ 3.3 &  $<$430   \\
& &	\multicolumn{4}{c}{(mJy km\,s$^{-1}$)} \\
\hline
B130\tablefootmark{*}  & 0.70  & 28 $\pm$ 6 & 116 $\pm$ 11 & 411 $\pm$ 82 & $<$36 \\
B380   & 1.13 & 50 $\pm$ 13 & 211 $\pm$ 21 & $<$53 & $<$69 \\
B630   & 3.24 & 306 $\pm$ 30 & 1060 $\pm$ 100 & $<$140 & $<$240 \\
B890   & 3.24 & 252 $\pm$ 33 & 956 $\pm$ 95 & $<$170 & $<$290 \\
B1130  & 2.26 & 163 $\pm$ 88 & 370 $\pm$ 110 & $<$350 & $<$500 \\
B1370  & 1.83 & $<$150 & 235 $\pm$ 73 & $<$180 & $<$330 \\
R130\tablefootmark{*}  & 0.85 & 89 $\pm$ 8 & 166 $\pm$ 16 & 223 $\pm$ 44 & $<$53 \\
R380   & 1.69 & 271 $\pm$ 27 & 671 $\pm$ 67 & $<$111 & $<$140 \\
R630   & 2.96 & 369 $\pm$ 36 & 1070 $\pm$ 110 & $<$190 & $<$240 \\
R890   & 3.39 & 258 $\pm$ 27 & 991 $\pm$ 99 & $<$190 & $<$240 \\
R1130  & 1.41 & $<$150 & 156 $\pm$ 50 & $<$200 & $<$260 \\
\hline
\end{tabular}
\end{small}
\tablefoot{The ``B'' and ``R'' regions correspond to the outflow annular sectors defined in Section~\ref{ss:outflow_sled} and shown in Figures~\ref{fig:annuli_blue} and \ref{fig:annuli_red}, where the number following the letter ``B'' or ``R'' indicates the projected distance to the nucleus in pc. The quoted errors include the statistical uncertainty, but not the flux accuracy uncertainties, which are 5\% for CO(1--0), 10\% CO(2--1), and 20\% CO(4--3) and CO(6--5).
\tablefoottext{a}{Total area where the CO emission is measured corrected for the inclination of the galaxy (i.e., observed area divided by $\cos 43\degree$ for the nucleus and the GMCs and by $\sin 43\degree$ for the outflow).}
\tablefoottext{b}{The nuclear fluxes are measured within a circular aperture of 2\farcs5 (580\,pc) diameter.}
\tablefoottext{*}{The CO outflow emission in the annulus sectors closer to the nucleus should be considered as a lower limit since the outflow is likely blended with the broad nuclear CO emission (see Section~\ref{ss:outflow_sled}).}}
\end{table*}

\begin{table*}[h]
\caption{Normalized CO SLEDs}
\label{tbl:sleds}
\centering
\begin{small}
\begin{tabular}{lcccccccc}
\hline \hline
\\
Region 			&  CO(1--0) & CO(2--1) & CO(4--3) & CO(6--5) \\
\hline
Nucleus 		&   0.24 $\pm$ 0.03  &  1.0 $\pm$ 0.1  & 2.9 $\pm$ 0.7  &  2.7 $\pm$ 0.6  \\
Disk GMC 		&   0.26 $\pm$ 0.05  &  1.0 $\pm$ 0.1  & 1.8 $\pm$ 0.5  &  $<1.4$    \\
Outflow ($d<250$\,pc)  	&   0.39 $\pm$ 0.15  &  1.0 $\pm$ 0.1  & 2.4 $\pm$ 1.0  &  $<1.0$    \\
Outflow ($d>250$\,pc) 	&   0.30 $\pm$ 0.10  &  1.0 $\pm$ 0.1  & $<0.6$ & $<0.8$     \\
\hline
\end{tabular}
\end{small}
\tablefoot{The CO SLEDs are normalized to the CO(2--1) flux. The ratios are calculated from the fluxes in Jy\,km\,s$^{-1}$. The quoted uncertainties include flux uncertainties and the standard deviation of the spatial variations within each region.}
\end{table*}

\begin{table*}[ht]
\caption{Bayesian inference results}
\label{tbl:bayes}
\centering
\begin{small}
\begin{tabular}{lcccccccccc}
\hline \hline
\\
Region & $T_{\rm kin}$ & $\log n_{\rm H_2}$ & $\log N_{\rm CO}\slash \Delta$v & $\log P$ & $\log \tau_{\rm 10}$ & $\log \Phi N_{\rm CO}$ & $\log \Phi \Delta$v & $\log \Phi R \frac{Z_{\rm CO}}{10^{-4}}$ & $\log \Delta$v$\slash r \frac{10^{-4}}{Z_{\rm CO}} $\\
& (K) & (cm$^{-3}$) & (cm$^{-2}$\slash km\,s$^{-1}$) & (K cm$^{-3}$) &  & (cm$^{-2}$) & (km\,s$^{-1}$) & (pc) &  (km\,s$^{-1}$\,pc$^{-1}$) \\
\hline
Nucleus & $37.6^{+119}_{-21.4}$ & $3.0^{+1.1}_{-0.7}$ & $17.7^{+1.1}_{-0.7}$ & $4.8^{+0.8}_{-0.9}$ & $1.5^{+1.8}_{-1.4}$ & $18.5^{+1.5}_{-0.7}$ & $0.7^{+0.4}_{-0.2}$ & $0.7^{+1.4}_{-1.3}$ & $0.0^{+1.3}_{-1.3}$ \\[0.3ex]
Disk GMC & $15.4^{+60}_{-6.8}$ & $3.0^{+1.1}_{-0.7}$ & $17.4^{+1.4}_{-0.9}$ & $4.4^{+0.9}_{-0.9}$ & $2.0^{+1.8}_{-2.1}$ & $17.5^{+1.8}_{-1.0}$ & $0.3^{+0.3}_{-0.3}$ & $-0.3^{+1.7}_{-1.4}$ & $0.5^{+1.5}_{-1.5}$ \\[0.3ex]
$\langle$Outflow$\rangle$\tablefootmark{a} & $9.4\pm1.2$ & $3.3\pm0.2$ & $16.2\pm0.4$ & $4.1\pm0.4$ & $0.5\pm1.1$ & $15.4\pm0.4$ & $-0.6\pm0.2$ & $-2.6\pm0.7$ & $1.8\pm0.7$ \\[0.3ex]
B130 & $15.6^{+28}_{-6.2}$ & $3.3^{+1.1}_{-0.9}$ & $17.5^{+1.2}_{-1.1}$ & $4.4^{+1.1}_{-0.9}$ & $1.8^{+1.7}_{-2.4}$ & $16.0^{+1.5}_{-1.2}$ & $-1.5^{+0.3}_{-0.3}$ & $-2.1^{+1.8}_{-1.8}$ & $0.5^{+1.8}_{-1.7}$ \\[0.3ex]
B380 & $11.2^{+23}_{-5.0}$ & $3.3^{+1.0}_{-0.8}$ & $16.0^{+1.3}_{-0.8}$ & $4.3^{+0.8}_{-1.1}$ & $-0.2^{+2.2}_{-0.8}$ & $15.0^{+1.4}_{-0.5}$ & $-0.8^{+0.4}_{-0.4}$ & $-3.1^{+2.0}_{-1.1}$ & $2.1^{+1.3}_{-2.0}$ \\[0.3ex]
B630 & $9.8^{+10}_{-2.9}$ & $3.4^{+0.9}_{-0.9}$ & $15.8^{+1.3}_{-0.6}$ & $4.5^{+0.7}_{-1.2}$ & $-0.2^{+1.5}_{-0.7}$ & $15.2^{+1.1}_{-0.4}$ & $-0.6^{+0.4}_{-0.3}$ & $-3.2^{+2.0}_{-0.9}$ & $2.5^{+1.1}_{-2.3}$ \\[0.3ex]
B890 & $11.1^{+14}_{-3.7}$ & $3.5^{+0.9}_{-0.8}$ & $15.7^{+1.3}_{-0.6}$ & $4.6^{+0.6}_{-1.2}$ & $-0.4^{+1.4}_{-0.6}$ & $15.0^{+1.0}_{-0.4}$ & $-0.7^{+0.4}_{-0.3}$ & $-3.3^{+1.9}_{-0.9}$ & $2.7^{+1.0}_{-2.1}$ \\[0.3ex]
B1130 & $8.9^{+40}_{-4.9}$ & $3.1^{+1.2}_{-0.8}$ & $16.7^{+1.9}_{-1.2}$ & $3.9^{+1.1}_{-0.9}$ & $2.1^{+2.2}_{-2.7}$ & $15.7^{+2.3}_{-1.0}$ & $-0.4^{+0.6}_{-0.7}$ & $-2.0^{+1.9}_{-1.5}$ & $1.2^{+1.5}_{-1.8}$ \\[0.3ex]
B1370 & $8.5^{+38}_{-4.5}$ & $3.1^{+1.2}_{-0.8}$ & $16.6^{+1.9}_{-1.2}$ & $3.9^{+1.1}_{-0.9}$ & $2.0^{+2.3}_{-2.6}$ & $15.6^{+2.3}_{-1.0}$ & $-0.5^{+0.6}_{-0.7}$ & $-2.1^{+1.9}_{-1.5}$ & $1.2^{+1.5}_{-1.7}$ \\[0.3ex]
R130 & $5.6^{+3.4}_{-1.4}$ & $3.8^{+0.8}_{-1.4}$ & $18.9^{+0.8}_{-1.8}$ & $4.4^{+1.0}_{-1.1}$ & $4.2^{+0.5}_{-1.1}$ & $18.3^{+1.0}_{-2.2}$ & $-0.5^{+0.5}_{-0.4}$ & $-0.5^{+1.4}_{-1.5}$ & $-0.1^{+1.4}_{-1.4}$ \\[0.3ex]
R380 & $5.9^{+12}_{-1.8}$ & $3.1^{+1.1}_{-0.8}$ & $16.7^{+1.3}_{-1.0}$ & $3.6^{+1.0}_{-0.7}$ & $1.9^{+1.9}_{-1.8}$ & $16.3^{+1.7}_{-1.0}$ & $-0.2^{+0.4}_{-0.4}$ & $-1.7^{+1.9}_{-1.4}$ & $1.2^{+1.5}_{-1.7}$ \\[0.3ex]
R630 & $7.9^{+14}_{-2.6}$ & $3.1^{+1.1}_{-0.8}$ & $16.5^{+1.1}_{-1.0}$ & $3.8^{+1.0}_{-0.8}$ & $1.0^{+1.6}_{-1.4}$ & $15.8^{+1.3}_{-0.7}$ & $-0.5^{+0.3}_{-0.3}$ & $-2.2^{+1.8}_{-1.2}$ & $1.5^{+1.5}_{-1.8}$ \\[0.3ex]
R890 & $12.1^{+16}_{-4.1}$ & $3.5^{+0.9}_{-0.7}$ & $15.6^{+1.1}_{-0.5}$ & $4.8^{+0.6}_{-1.1}$ & $-0.5^{+0.8}_{-0.5}$ & $14.9^{+0.8}_{-0.3}$ & $-0.7^{+0.4}_{-0.4}$ & $-3.5^{+1.3}_{-0.8}$ & $2.8^{+1.0}_{-1.6}$ \\[0.3ex]
R1130 & $9.5^{+42}_{-5.4}$ & $3.1^{+1.2}_{-0.8}$ & $16.7^{+1.9}_{-1.2}$ & $3.9^{+1.2}_{-0.9}$ & $2.1^{+2.2}_{-2.6}$ & $15.6^{+2.2}_{-1.0}$ & $-0.5^{+0.6}_{-0.8}$ & $-2.2^{+1.9}_{-1.6}$ & $1.1^{+1.5}_{-1.8}$ \\[0.3ex]
\hline
\end{tabular}
\end{small}
\tablefoot{The quoted errors correspond to the $\pm$1$\sigma$ range from the Bayesian inference.
\tablefoottext{a}{Average values calculated for the outflowing gas at radial distances from the nucleus $>250$\,pc (i.e., excluding the B130 and R130 annulus sectors; see also Figure~\ref{fig:radial}).}
}
\end{table*}

\subsection{Nuclear, disk, and outflow CO SLED comparison}\label{ss:CO_SLEDS}

We also measured the CO SLED of the nucleus using a $r\sim$1\arcsec (230\,pc) aperture. For the disk, we estimated the average CO SLED produced in giant molecular clouds (GMCs). To do so, we defined 126 nonoverlapping regions in the disk, with radius of about 0\farcs3 (70\,pc), from the emission peaks detected at $>$3$\sigma$ in the CO(2--1) moment 0 map (second panel of Figure~\ref{fig:moment0}). 
Then, we normalized the SLED to the CO(2--1) flux and calculated the mean value for each transition to obtain the average CO SLED.
We followed the same procedure to calculate the average SLED for the each of the outflow sectors defined in Section~\ref{ss:outflow_sled} (see also Figures~\ref{fig:annuli_blue} and \ref{fig:annuli_red}).

The total measured CO fluxes in the nucleus, the disk GMCs, and the outflow annular sectors, are presented in Table~\ref{tbl:co_fluxes}. The normalized SLEDs are plotted in Figure~\ref{fig:co_sled} and listed in Table\,\ref{tbl:sleds}. In this figure, we only consider 2 average SLEDs for the outflowing gas for radial distances lower and greater than $\sim$250\,pc. {This is done to distinguish between the outer outflow regions, where the CO(4--3) transition is undetected, and the innermost annular sector of the outflow, where the CO(4--3) transition is detected but its flux is uncertain due to broad nuclear emission profile (see Figures~\ref{fig:annuli_blue} and \ref{fig:annuli_red}).}
Figure~\ref{fig:co_sled} shows that the CO in the nucleus is more excited than in the disk and the outflow. The CO(6--5) transition is only detected in the nucleus (nuclear CO(6--5)\slash CO(2--1) ratio equals to 2.7$\pm$0.6) while the CO(6--5) upper limits in the disk and outflow indicate CO(6--5)\slash CO(2--1) ratios lower than 1.4--0.8. Similarly, the nuclear CO(4--3)\slash CO(2--1) ratio (2.9$\pm$0.7) is higher than the disk (1.8$\pm$0.5) and outer outflow ($<0.6$) ratios. 
Finally, the nuclear CO(1--0)\slash CO(2--1) ratio is slightly lower than the ratio measured in the disk and outflow, which is consistent with the enhanced excitation of the nuclear molecular gas too.

This comparison also shows that the excitation of molecular gas in the outer outflow is reduced with respect to that of the nucleus and the disk GMCs. For instance, the CO(4--3)\slash CO(2--1) upper limit for the outer outflow is $<$0.8, while in the disk the average CO(4--3)\slash CO(2--1) ratio is 1.8$\pm$0.5.
This lower excitation in the outer outflow with respect to the disk is evident in the spectra of regions 2 and 4 shown in Figures~\ref{fig:annuli_blue} and \ref{fig:annuli_red}, respectively, where the CO(4--3) narrow disk component (i.e., GMC emission) is clearly detected but the high-velocity CO(4--3) emission is below the detection limit. {This confirms the lower excitation of the outflowing molecular gas compared to that of the molecular gas in the nuclear region where the outflow originates.}
Only the most central ($r_{\rm proj}<250$\,pc) outflowing gas seems to have a CO(4--3)\slash CO(2--1) ratio {compatible with} that measured in the nucleus. Although the upper limit for the CO(6--5)\slash CO(2--1) ratio ($<$1.0) in this inner outflow is well below the ratio observed in the nucleus (3.9$\pm$0.9). {However, we note that the comparison between the innermost outflow region and nuclear emission is uncertain since the broad nuclear CO profiles might contaminate the outflow flux measurement (see Section~\ref{ss:outflow_sled} and Figures~\ref{fig:annuli_blue} and \ref{fig:annuli_red}).}

\begin{figure*}
\centering
\vspace{5mm}
\includegraphics[width=0.82\textwidth]{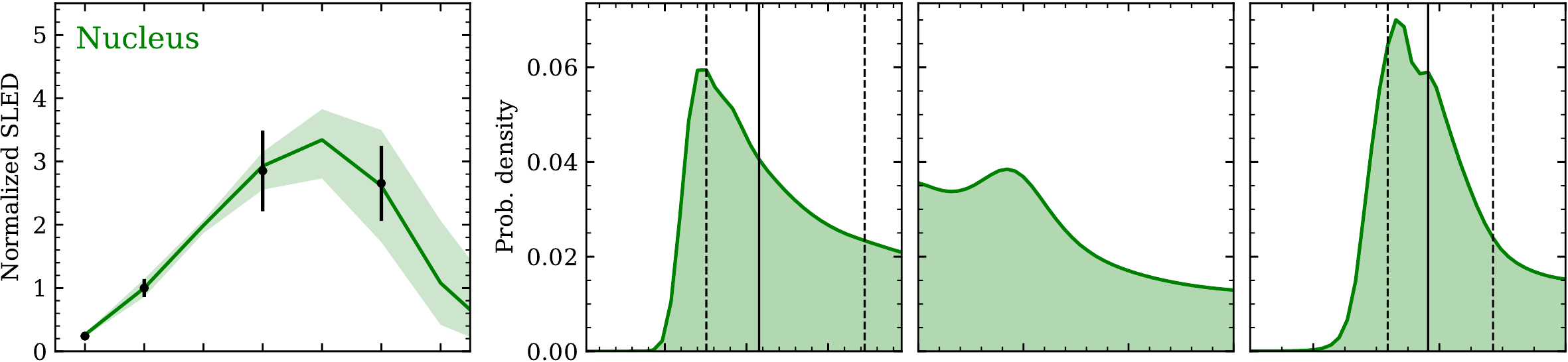}
\includegraphics[width=0.82\textwidth]{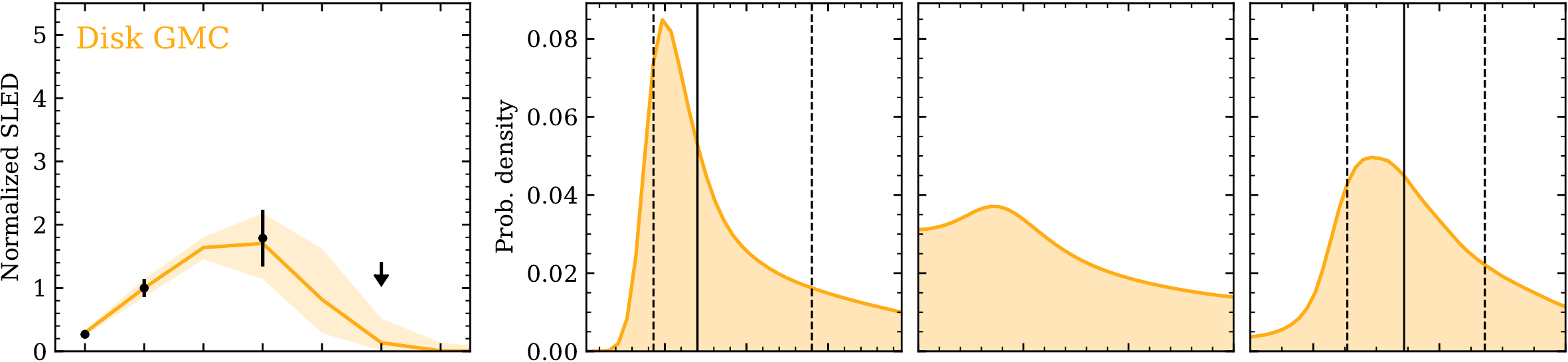}
\includegraphics[width=0.82\textwidth]{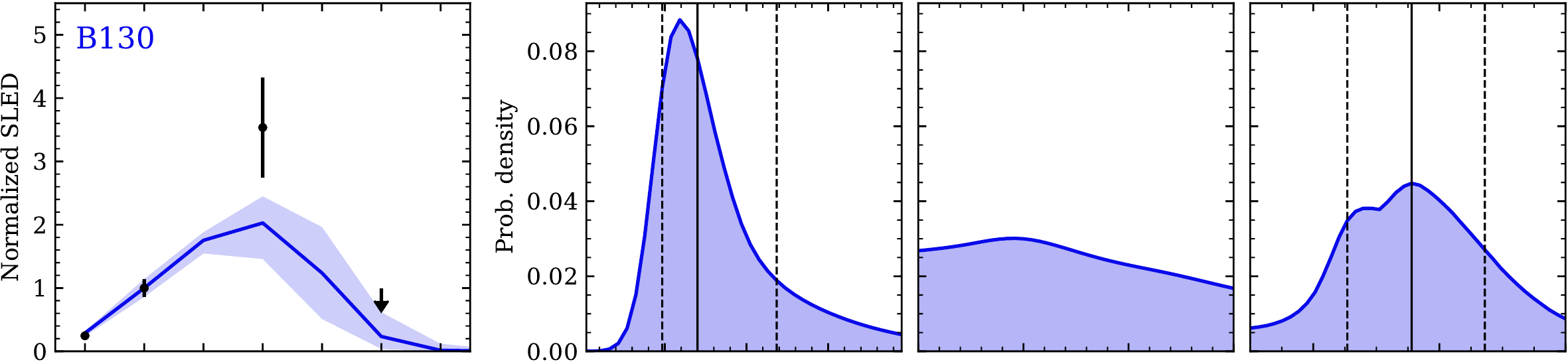}
\includegraphics[width=0.82\textwidth]{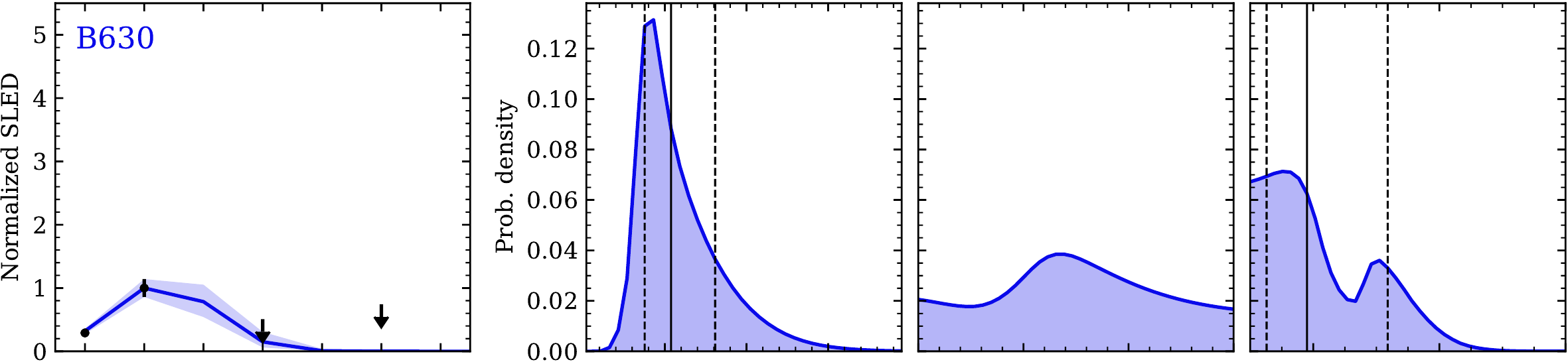}
\includegraphics[width=0.82\textwidth]{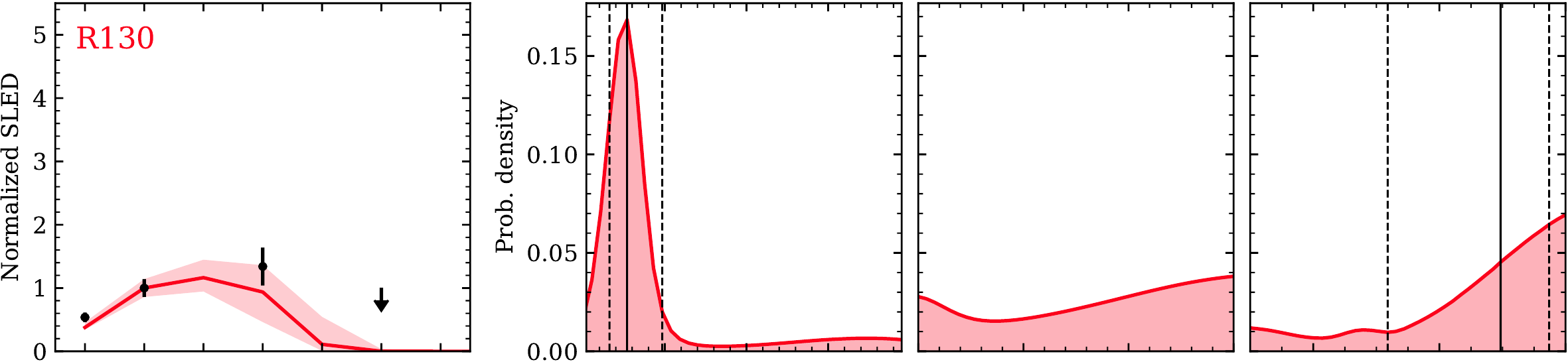}
\includegraphics[width=0.82\textwidth]{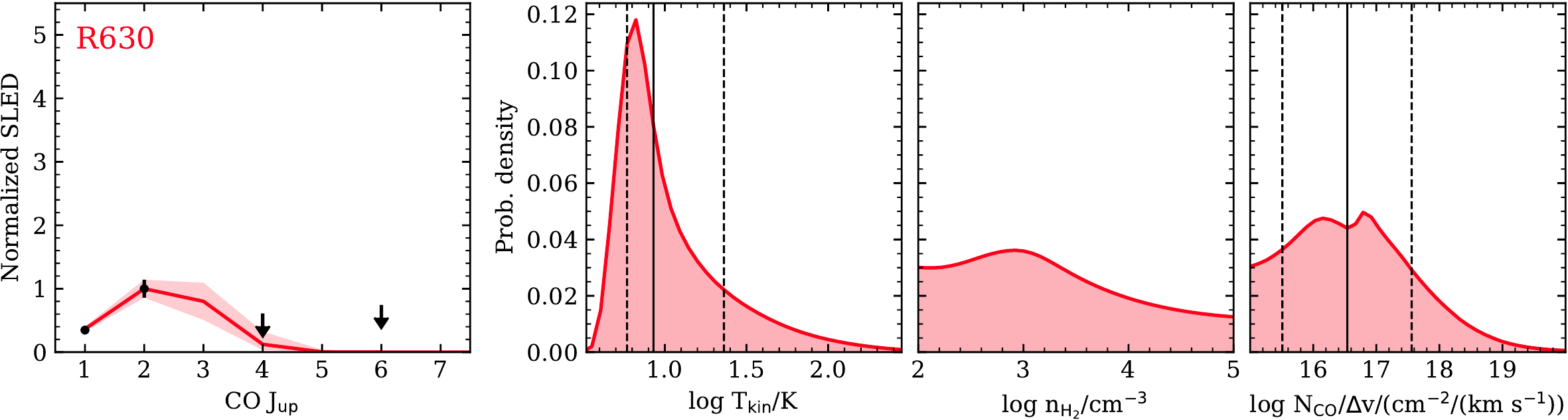}
\caption{Row show the best fitting CO SLEDs models and the probability density distributions of the model parameters ($T_{\rm kin}$, $n_{\rm H_2}$, and $N_{\rm CO}\slash \Delta$v) for the nucleus, the average disk GMC, and four annulus sectors of the blue- and red-shifted outflow emission (B130, B630, R130, and R630). The left panels show the normalized CO SLED (black points and upper limits) and the best fitting SLED (color solid line) and its 1$\sigma$ dispersion (color shaded area). The other panels show the probability density distributions. The vertical solid line marks the median value and the dashed lines the $\pm$1$\sigma$ range. We do not show the median and $\pm$1$\sigma$ range in the $n_{\rm H_2}$ panel because it is not well constrained by these observations. These distributions are obtained assuming a uniform prior for the three parameters.
\label{fig:bayes_1}}
\end{figure*}

\subsection{Non-LTE CO radiative transfer models}\label{ss:radex_models}

To analyze the CO SLEDs, we generated a grid of non-LTE radiative transfer models using the \textsc{RADEX} software \citep{vanderTak2007}. \textsc{RADEX} uses the escape probability approximation to solve the radiative transfer equations and the molecular level populations, and then it calculates the intensities of the emission lines and their optical depths ($\tau$).

The \textsc{RADEX} physical input parameters are the kinetic temperature ($T_{\rm kin}$), the H$_2$ number density ($n_{\rm H_2}$), and the ratio between the CO column density and the line width ($N_{\rm CO}\slash \Delta$v). We include collisions with ortho- and para-H$_2$ using the \citet{Yang2010} collisional coefficients and assuming the LTE H$_2$ ortho-to-para ratio, which varies between 0 for low temperature and 3 for $T_{\rm kin} > $200\,K.

In our grid, we cover a wide range for these parameters but we also try to limit the number of models which produce the same line ratios (e.g., gas in LTE due to a high number density or very optically thick transitions) to avoid biasing the subsequent the Bayesian inference.
We consider $T_{\rm kin}$ between 3 and 400\,K, $n_{\rm H_2}$ between 10$^2$ and 10$^5$\,cm$^{-3}$, and $N_{\rm CO}\slash \Delta$v from 10$^{15}$ to 10$^{20}$ cm$^{-2}$\slash km\,s$^{-1}$. The grid consists of 40 steps for each parameter in log space and, therefore, a total of 64000 models.
The temperature range is comparable to the upper levels energies of the CO transitions discussed in this paper (from 5 to 115\,K). 
For the H$_2$ number density, the upper end of the range is set by the critical density of these transitions. When the H$_2$ density is above the critical density (10$^{3.5}$--10$^{5.5}$\,cm$^{-3}$ for these transitions), the level populations are thermalized (i.e, LTE) and the modeled emission line ratios are not sensitive to density variations.
The CO column density range is such that it covers from optically thin emission ($\tau \ll $1) to  optically thick ($\tau \sim $10--100). Therefore, beyond this range, it becomes impossible to use these CO line ratios to estimate the column density.

\subsection{Bayesian CO SLED parameter inference}\label{ss:model_results}

\begin{figure*}
\centering
\vspace{5mm}
\includegraphics[width=0.3\textwidth]{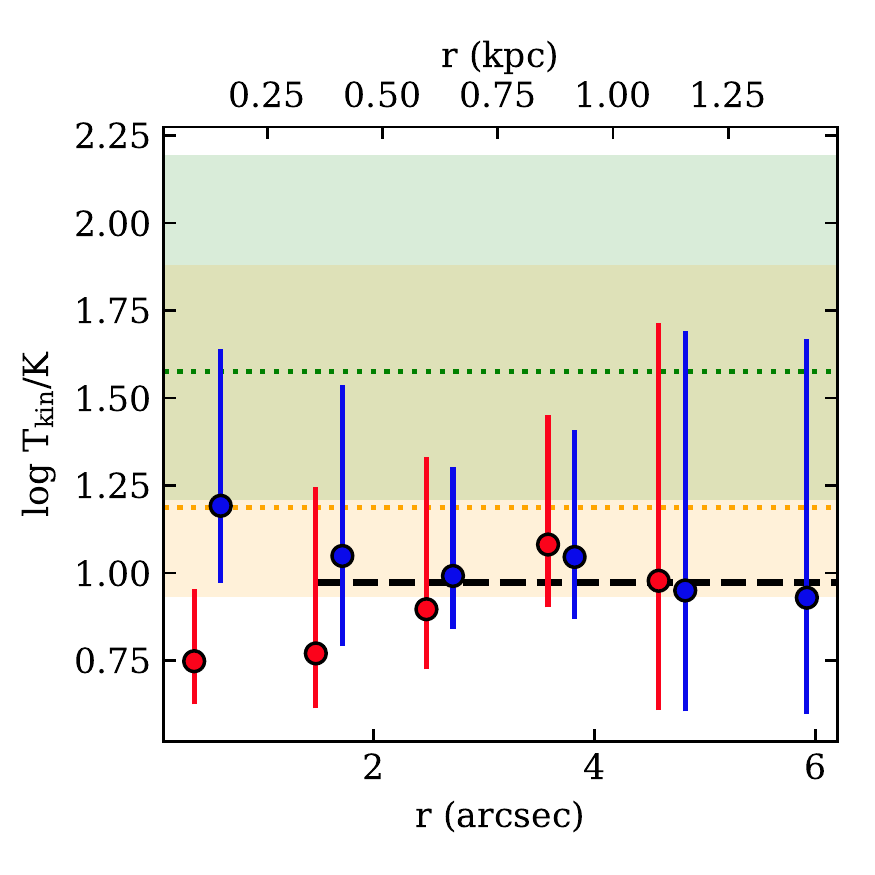}
\includegraphics[width=0.3\textwidth]{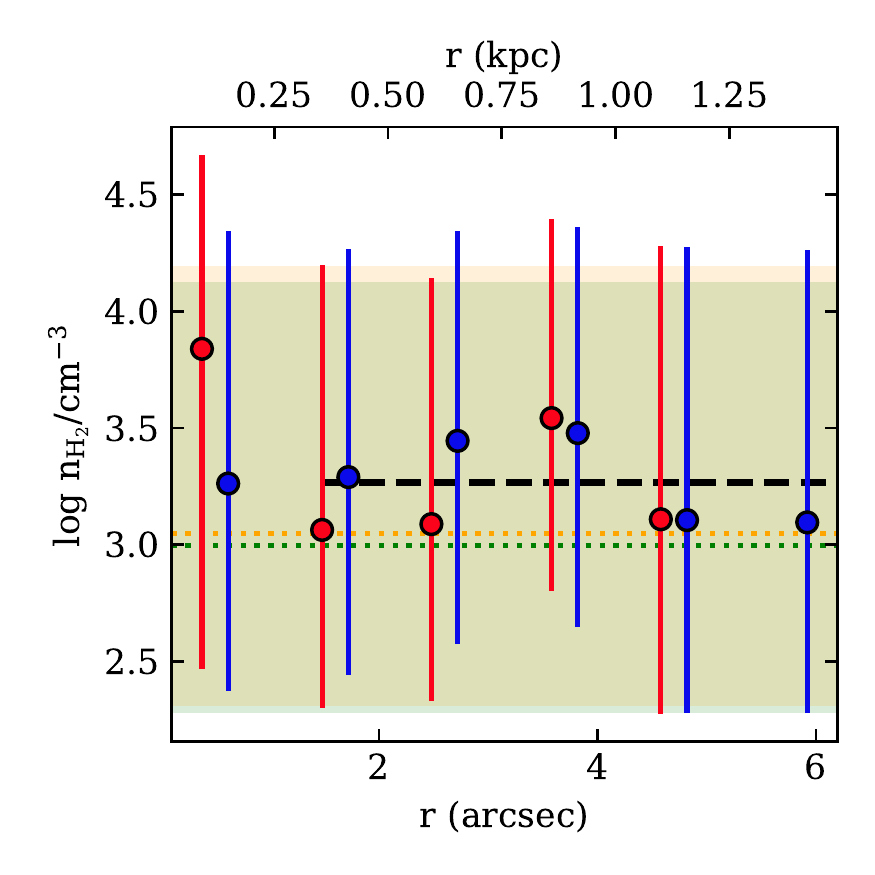}
\includegraphics[width=0.3\textwidth]{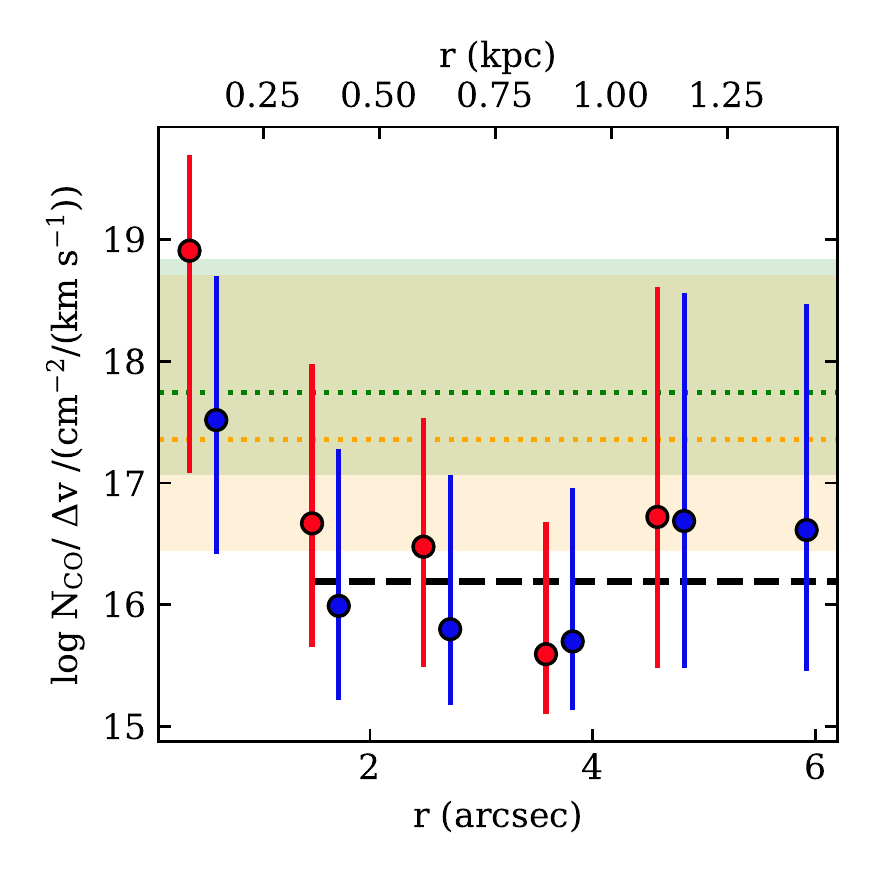}
\includegraphics[width=0.3\textwidth]{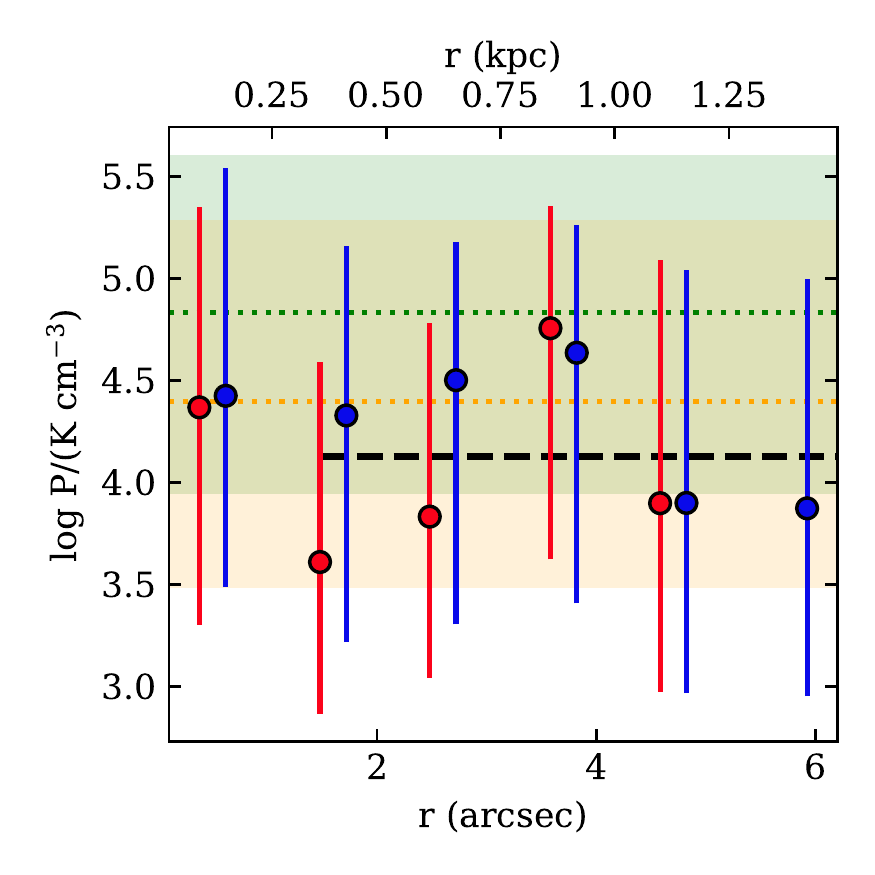}
\includegraphics[width=0.3\textwidth]{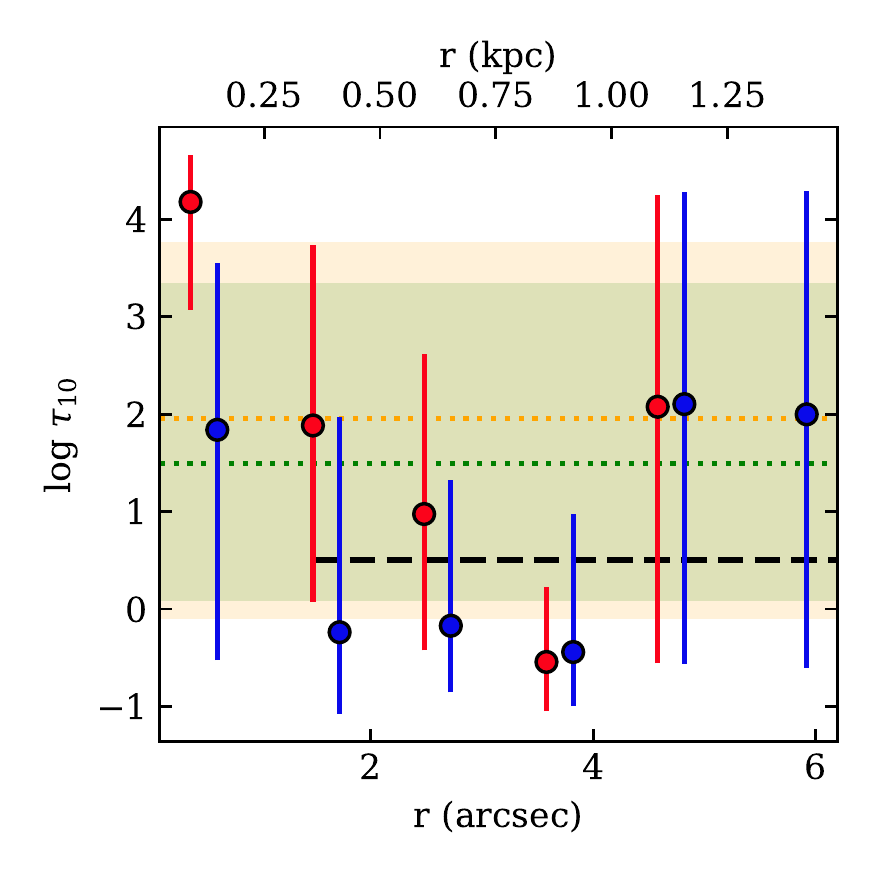}
\includegraphics[width=0.3\textwidth]{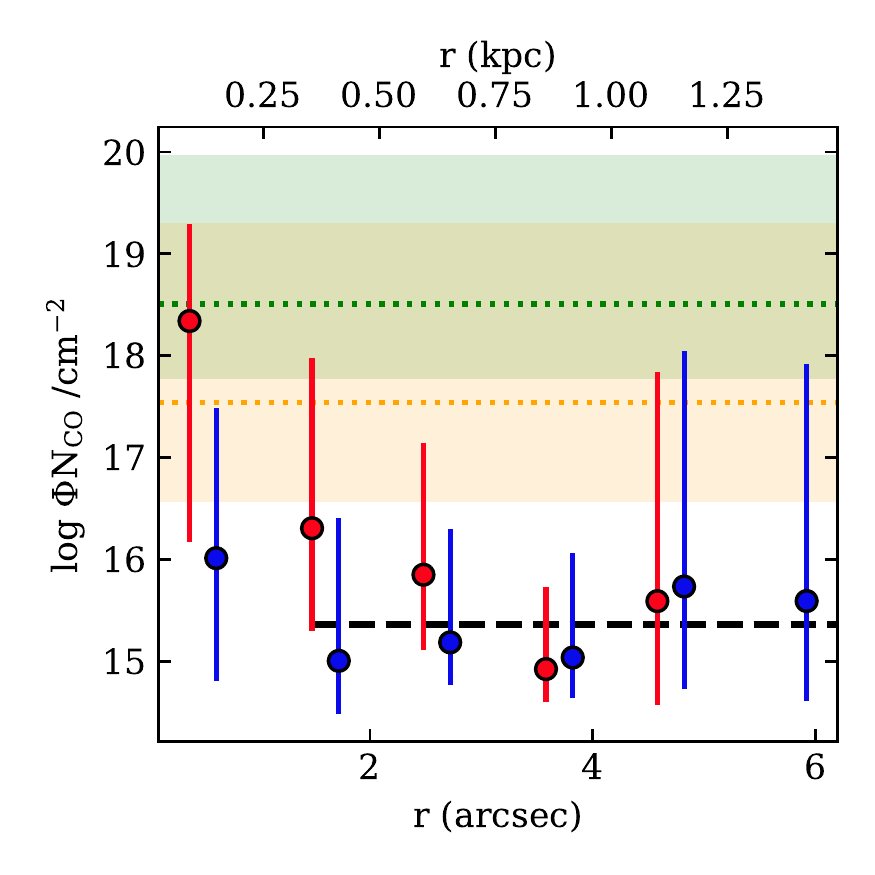}
\includegraphics[width=0.3\textwidth]{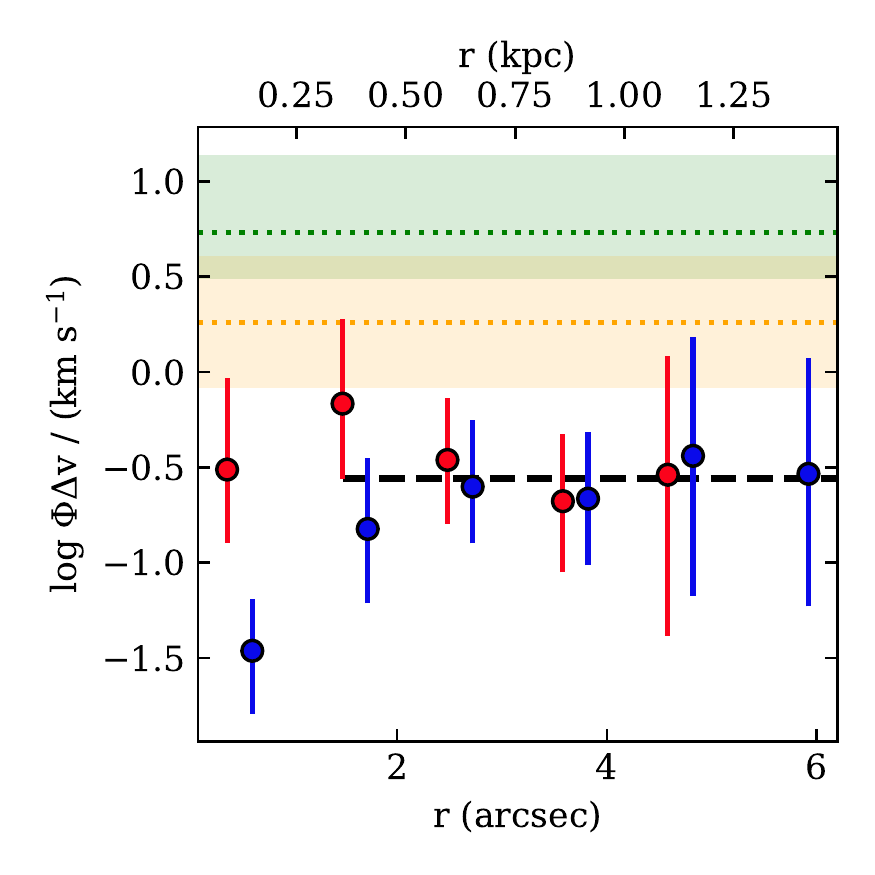}
\includegraphics[width=0.3\textwidth]{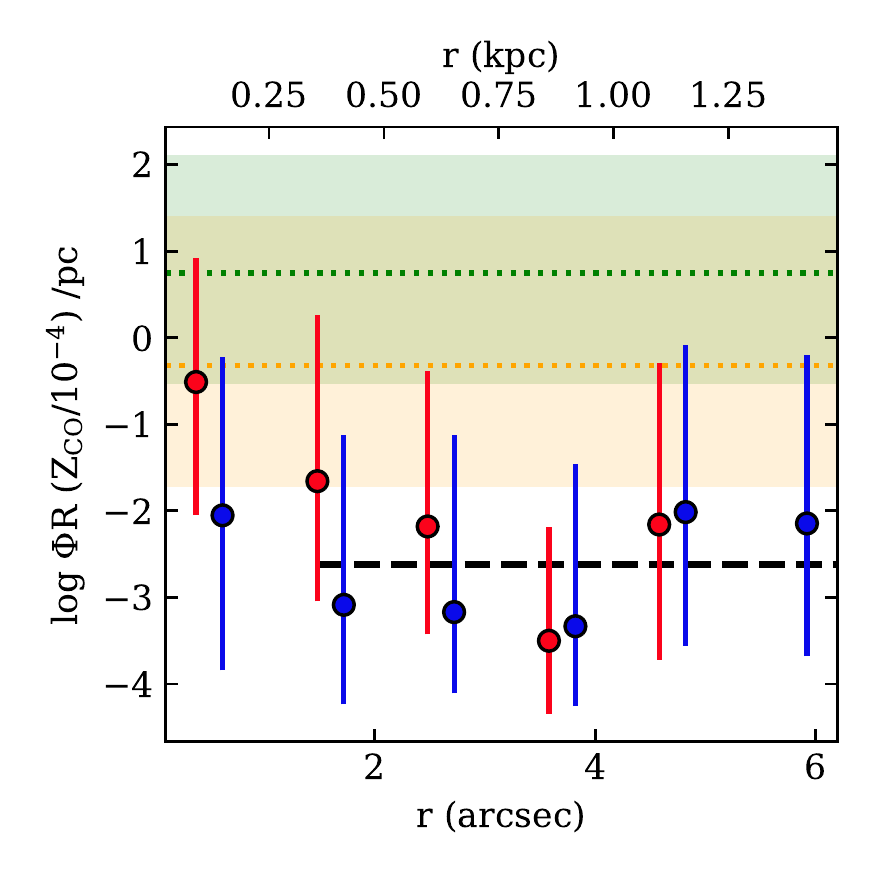}
\includegraphics[width=0.3\textwidth]{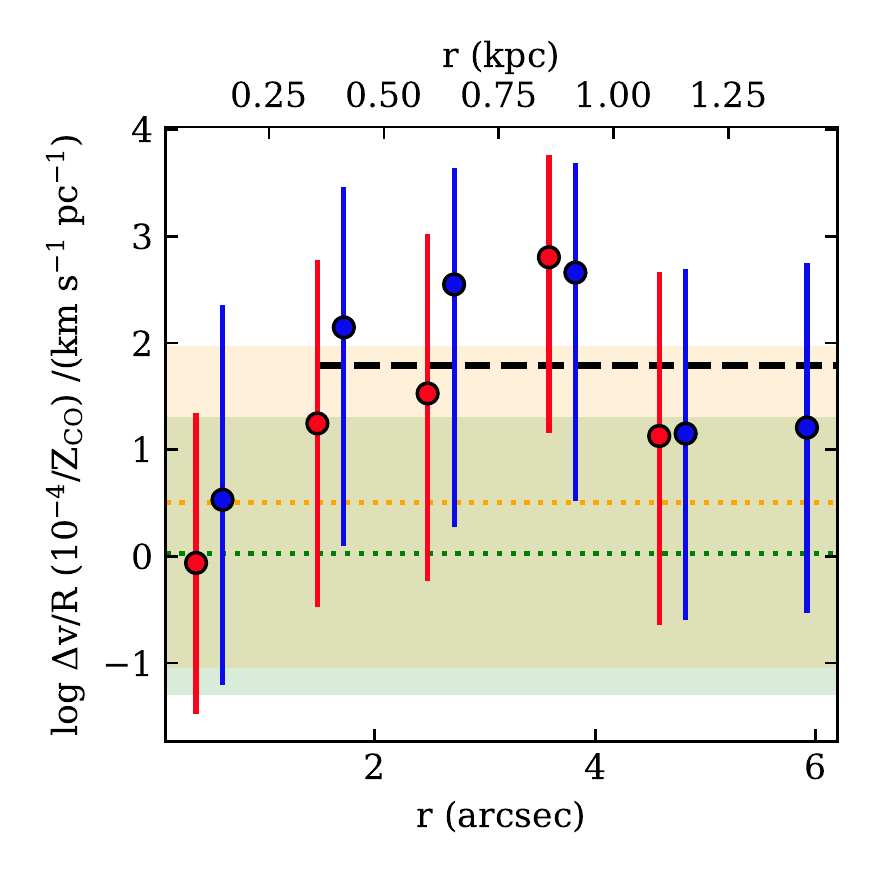}
\caption{Radial dependence of the parameters (kinetic temperature $T_{\rm kin}$, number density $n_{\rm H_2}$, ratio between the CO column density and the line width $N_{\rm CO}\slash \Delta$v, pressure $P$, optical depth of the CO(1--0) transition $\tau_{10}$, angular area filling factor $\Phi$ times $N_{\rm CO}$, $\Delta$v, and the radius of the cloud $R$, and the $\Delta$v\slash $R$ ratio) derived from the radiative transfer modeling (Section~\ref{ss:radex_models} and Table~\ref{tbl:bayes}). The red and blue points correspond to the red- and blue-shifted sides of the outflow.
The dashed black line is the average outflow value for $r_{\rm proj}>250$\,pc. The green (orange) shaded range and the dashed green (orange) line are the 1$\sigma$ range and average value for the nucleus (disk GMC). 
The dashed black line corresponds to the average value in the outflow for $r>250$\,pc.
\label{fig:radial}}
\end{figure*}

We used the models described in the previous section to investigate the physical conditions of the molecular gas from the CO SLEDs (Section~\ref{ss:CO_SLEDS}). 
For each CO SLED, we calculated the $\chi^2$ for the 64000 models and assigned a probability to each model proportional to $\exp(-\chi^2\slash 2)$. To take into account the upper limits of the SLEDs, we multiplied this probability by $1 + {\rm erf}((L - M)\slash (\sqrt{2}\,L))$, where $L$ is the 1$\sigma$ upper limit and $M$ the prediction of the \textsc{RADEX} model (see e.g., Appendix B of \citealt{Pereira2015not}). This is equivalent to assuming a uniform prior for the probabilities of the model parameters in their given ranges. 
Then, from these probabilities, we estimated the marginalized probability density function for the model parameters. In addition, we calculated the probability densities for other physical quantities derived from the models like the pressure ($P=T_{\rm kin}\times n_{\rm H_2}$), the optical depth of the transitions, and the normalized CO SLED. 
We also considered the angular area filling factor $\Phi$ (i.e., the ratio between the observed surface brightness and that predicted by \textsc{RADEX}). We did not fix $\Delta$v, so the surface brightness predicted by the models depends on $\Delta$v and the $N_{\rm CO}\slash\Delta$v ratio. Therefore, we cannot estimate the probability density for $\Phi$. Instead, we can determine the probability densities for $\Phi N_{\rm CO}$ and $\Phi \Delta$v. We can also estimate the $\Phi R$ probabilities density, where $R$ is the radius of the clouds defined as $R = 0.5\times N_{\rm H_2}\slash n_{\rm H_2} = N_{\rm CO}\slash Z_{\rm CO}\slash n_{\rm H_2}$, where $Z_{\rm CO}$ is the CO to H$_2$ abundance ratio. We assumed a fiducial $Z_{\rm CO}=10^{-4}$. Finally, we also calculate the probability density for the velocity gradient of the clouds, $\Delta$v$\slash R$.
We note that the latter does depend on the CO abundance, $Z_{\rm CO}$, but not on the filling factor, $\Phi$.
The parameters chosen to create the grid of models (i.e., $T_{\rm kin}$, $n_{\rm H_2}$, and $N_{\rm CO}\slash \Delta$v) are uniformly sampled in log space. However, this implies that the values of the derived quantities are not sampled uniformly. 
Therefore, we weighted the probability of the models to obtain a uniform prior for each of these derived quantities.

We present the result of the Bayesian inference in Figure~\ref{fig:bayes_1} and Table\,\ref{tbl:bayes}. The best-fitting SLEDs (left panels of Figure~\ref{fig:bayes_1}) show that the models properly reproduce the observed SLEDs, except the CO(4--3) transition measured in the innermost outflow annulus sectors (B130 and R130). These two outflow SLEDs present a CO(4--3) emission excess that cannot be explained by the adopted models. This could indicate that: the inner high-velocity molecular gas has various phases with different kinetic temperatures and densities; or that the CO(4--3) measurements are contaminated by the bright nuclear CO(4--3) emission. To investigate these options higher spatial resolution data would be required.

\begin{figure}
\centering
\includegraphics[width=0.38\textwidth]{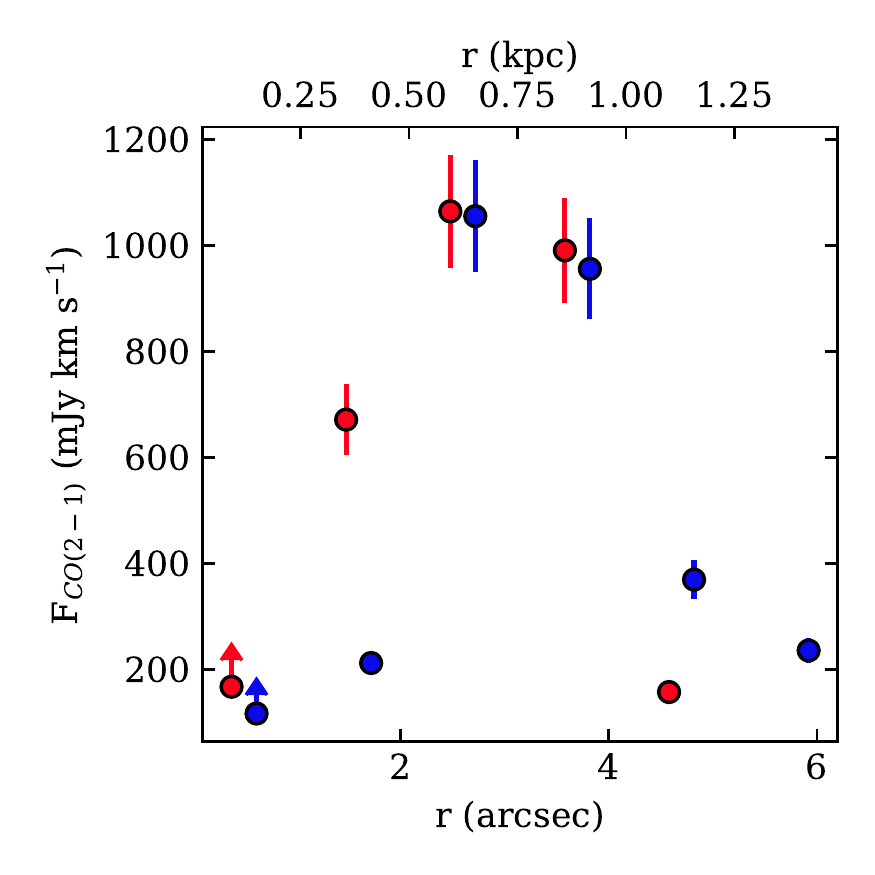}
\caption{CO(2--1) flux of the outflow as a function of the projected distance to the nucleus measured in the annulus sector. The red and blue points correspond to the red- and blue-shifted sides of the outflow. The fluxes of the inner annulus sector ($r_{\rm proj}\sim130$\,pc) should be considered as lower limits (see Section~\ref{ss:outflow_sled}).\label{fig:radial_flux}}
\end{figure}

In Table~\ref{tbl:bayes}, we also include the averaged value of the outflow parameters for regions with distances $>250$\,pc. This spatial average is motivated because the radial variation of the parameters beyond $r_{\rm proj}>250$\,pc is small, as shown in Figure~\ref{fig:radial}, and also because most of the high-velocity molecular emission is detected at $r_{\rm proj}>250$\,pc (see Figure~\ref{fig:radial_flux}). 
The innermost annulus sector ($r_{\rm proj}\sim$130\,pc) is excluded from this average because its column density and optical depth are higher than in the rest of the outflow (see Table~\ref{tbl:co_fluxes} and Figure~\ref{fig:radial_flux}) and the outflow conditions appear to be different closer to the launching region.

\subsection{Summary of the CO SLED parameter inference}

Figure~\ref{fig:radial} summarizes the results of the non-LTE CO radiative transfer analysis. The nuclear $T_{\rm kin}$ is higher than that of the disk GMCs (38$^{+119}_{-21}$ vs. 15.4$^{+60}_{-6.8}$\,K) and these two are higher than the average outflow $T_{\rm kin}$ (8.9$\pm$1.4\,K).
Also, the models show that the $N_{\rm CO}\slash \Delta$v is slightly reduced in the outflow with respect to that of the nucleus and the disk. The H$_2$ number density and the optical depth are not well constrained by these data, although their probability density distributions peak around 10$^{3-3.5}$\,cm$^{-3}$ and 10$^{0-2}$, respectively. The pressure is also similar within the uncertainties, around 10$^{4.2}$--10$^{4.8}$\,K\,cm$^{-3}$, in the nucleus, disk, and outflow.
Finally, the panels with the angular area filling factor $\Phi$ present the largest differences between the gas in the outflow and in the rest of the galaxy. The outflow $\Phi N_{\rm CO}$ and $\Phi \Delta$v are lower than in the disk and nucleus. The $\Phi R$ distributions suggest that the outflow clouds are smaller than those in the nucleus and disk. Although, if the CO abundance or the filling factor in the outflow are lower, the clouds could have larger sizes.

In summary, the molecular gas in this outflow seems to be colder (lower $T_{\rm kin}$), distributed in lower column density clouds (lower $N_{\rm CO}\slash \Delta$v) than the nucleus, where the outflow is produced, and the disk of the galaxy. 
The filling factor panels in Figure~\ref{fig:radial} suggest that the outflowing clouds are either smaller, have lower CO abundances, or lower filling factors (e.g., from the $\Phi R$ probability density). The $\Delta$v$\slash R$ ratio in the outflow indicates that the outflowing clouds have larger velocity gradients, assuming similar CO abundances, and, therefore, they are farther from virial equilibrium than disk and nuclear clouds. 
Consequently, the physical processes that control the heating\slash cooling, stability, collapse and chemical composition of molecular clouds in the outflow and in the disk of this object are likely to be different.

\section{Evolution of the outflow molecular phase}\label{s:evol}

In this section, we discuss the thermal and dynamical evolution of the outflow based on the spatially resolved multi-transition CO ALMA observations of the ESO~320-G030 outflow.

\subsection{Thermal evolution}\label{ss:thermal_ev}

The origin of the molecular gas in the outflow is not well established. It might condensate in the outflow (e.g., \citealt{Ferrara2016}), it might be molecular gas from the galaxy entrained and accelerated (e.g., \citealt{McCourt2015}), or a combination of both mechanisms (e.g., \citealt{Gronke2020}). To investigate its origin, we discuss its thermal evolution in this section.

The heating and cooling of the cold gas in outflows are important processes that affect the evolution of these cold clouds surrounded by a hot and low-density medium. According to hydrodynamical simulations, when the cooling of the gas is efficient, the evaporation rate of the cold clouds is reduced since the development of hydrodynamical instabilities is hindered (e.g., \citealt{Bruggen2016, Armillotta2017}). However, we also note that the ``cold'' gas in most hydrodynamical models has $T\sim10^4$\,K and is in atomic phase, so this could be different for the molecular clouds discussed here.

We use the results from the radiative transfer models (Section~\ref{ss:model_results}) to determine the cooling and heating of the outflowing molecular gas. The outflow originates at the compact starburst in the nucleus of ESO~320-G030. The kinetic temperature of the molecular gas in the nucleus is 38$^{+119}_{-21}$\,K while the outflowing molecular gas has a lower average temperature of 8.9$\pm$1.4\,K (Table~\ref{tbl:bayes}). So if the outflowing molecular gas was entrained and accelerated it must have cooled down in timescales shorter than the flow time of the outflowing gas that is detected closer to the nucleus, $\sim$1.5\,Myr $=$ 360\,pc\slash 250\,km\,s$^{-1}$\slash $\tan i$, where $i$ is the galaxy inclination (assuming an outflow perpendicular to the disk; see also Section~\ref{ss:dynamical_ev}). We approximate an upper limit on the cooling time $t_{\rm cool} < \left(3/2 k \Delta T\right)$\slash $\Lambda_{\rm 10\,K}\sim 100$\,kyr, assuming a conservative cooling rate $\Lambda_{\rm 10\,K}$ of 10$^{-26.7}$\,erg\,s$^{-1}$ per H$_2$ molecule for molecular gas at 10\,K \citep{Neufeld1995}, and a temperature variation, $\Delta T$, of 30\,K. The cooling rate at 38\,K, $\Lambda_{\rm 38\,K}$, is about 10 times higher than $\Lambda_{\rm 10\,K}$, so the nuclear gas at 38\,K would reach the $\sim$9\,K outflow temperature even in shorter timescales.
Therefore, if the outflowing molecular gas is molecular gas accelerated from the launch site, it should be possible for this hotter nuclear molecular gas to cool down while it is moving away from the nucleus.

We also find that the kinetic temperature in the outflow is relatively constant (first panel of Figure~\ref{fig:radial}). This suggests that the molecular gas is at thermal equilibrium, so the heating by the external hot medium would be similar to the cooling of the cold molecular gas $\Lambda$. This low equilibrium temperature of $\sim$10\,K compared to the $>$10$^4$\,K and $\sim$10$^7$\,K of the ionized and X-ray plasma phases of the outflow, respectively, suggests that the heating is not very efficient and this could enhance the survival of the molecular gas in the outflow.

For the outflow, the kinetic temperature obtained from the \textsc{RADEX} model is determined by the \hbox{CO(2--1)\slash CO(1--0)} ratio, which is lower in the outflow than in the disk and the nucleus, and the upper limit on the CO(4--3) outflow emission (see Table~\ref{tbl:sleds} and Figure~\ref{fig:co_sled}). A reduced CO(2--1)\slash CO(1--0) ratio with respect to the disk is also observed in the outflows of other starburst galaxies (M82, \citealt{Weiss2005}; NGC~253, \citealt{Zschaechner2018}; NGC~1808 shows a reduced CO(3--2)\slash CO(1--0) ratio in the outflow \citealt{Salak2018}). So, these observations support that the molecular gas in starburst-driven outflows is less excited than the gas in the disk.
However, in jet-driven outflows the behavior of the CO line ratios seems to be the opposite. For instance, the CO(2--1)\slash CO(1--0) ratio is increased in the high-velocity outflow emission of the early-type AGN NGC~1266 \citep{Alatalo2011}. Also, the radio-loud Seyfert IC~5063 has enhanced \hbox{CO(4--3)\slash CO(2--1)} ratios in the outflow \citep{Dasyra2016} and, similarly, the AGN-driven outflows of NGC~1068 \citep{GarciaBurillo2014, Viti2014} and the southern nucleus of LIRG NGC~3256 show higher \hbox{CO(3--2)\slash CO(1--0)} ratios in the outflow than in the nuclear regions too \citep{Michiyama2018}. However, in the ULIRG quasar Mrk~231, the CO(2--1)\slash CO(1--0) is similar in the broad and narrow components \citep{Cicone2012}. Therefore, the physical conditions of the outflowing molecular gas seem to be different in starburst- and most of the AGN-driven outflows.

\subsection{Outflowing clouds stability}\label{ss:statbility}

We compare the estimated velocity gradient from the LVG model ($\Delta$v$\slash R$ in Figure~\ref{fig:radial} and Table~\ref{tbl:bayes}) with the virial velocity gradient $(\Delta$v$\slash R)_{\rm vir}$ to study the cloud stability. We use the parameter $K_{\rm vir}$ defined as:
\begin{equation}
K_{\rm vir} = \frac{(\Delta {\rm v}\slash R)_{\rm LVG}}{(\Delta {\rm v}\slash R)_{\rm vir}}.
\end{equation}
$K_{\rm vir}\sim 1$ indicates that the cloud is virialized while $K_{\rm vir}> 1$ indicates that the gas is not bound by self-gravity.
The virial velocity gradient is $(\Delta$v$\slash R)_{\rm vir}$:
\begin{equation}
(\Delta {\rm v}\slash R)_{\rm vir} \sim 0.65 \alpha^{1\slash 2}\left( \frac{\langle n_{\rm H_2} \rangle}{10^3\,{\rm cm^{-3}}} \right)^{1\slash 2},
\end{equation}
where $\alpha$ is between 0.6 and 2.4 and depends on the cloud density distribution \citep{Bryant1996}, and $\langle n_{\rm H_2} \rangle$ is the mean number density of the cloud (see \citealt{Papadopoulos1999}). For the densities favored by our models, {10$^{2.0}$ to 10$^{4.0}$\,cm$^{-3}$}, and considering the whole range of $\alpha$, $(\Delta$v$\slash R)_{\rm vir}$ ranges from {0.05 to 10}\,km\,s$^{-1}$\,pc$^{-1}$.

Assuming the fiducial CO abundance $Z_{\rm CO}$ of 10$^{-4}$, the \hbox{$(\Delta$v$\slash R)_{\rm LVG}$} {ratio} of the nucleus and the disk are 1 and 3\,km\,s$^{-1}$\,pc$^{-1}$, respectively. The uncertainties are large, about $\pm$1.4\,dex, but these values are compatible with self-gravity bound clouds {($K_{\rm vir}$=0.1--60)}. For the outflow, \hbox{$(\Delta$v$\slash R)_{\rm LVG}$}=60$^{+250}_{-45}$\,km\,s$^{-1}$\,pc$^{-1}$.
This corresponds to {$K_{\rm vir}$=2--6000}, so it suggests that the outflowing clouds are not gravitationally bound, unless the CO abundance is {at least $>$2} times lower in the outflow than the assumed $Z_{\rm CO}$ = 10$^{-4}$. Although, the clouds could survive if they are confined by an external pressure (e.g., \citealt{Field2011}).

\subsection{Dynamical evolution}\label{ss:dynamical_ev}

\begin{figure*}
\centering
\includegraphics[width=0.75\textwidth]{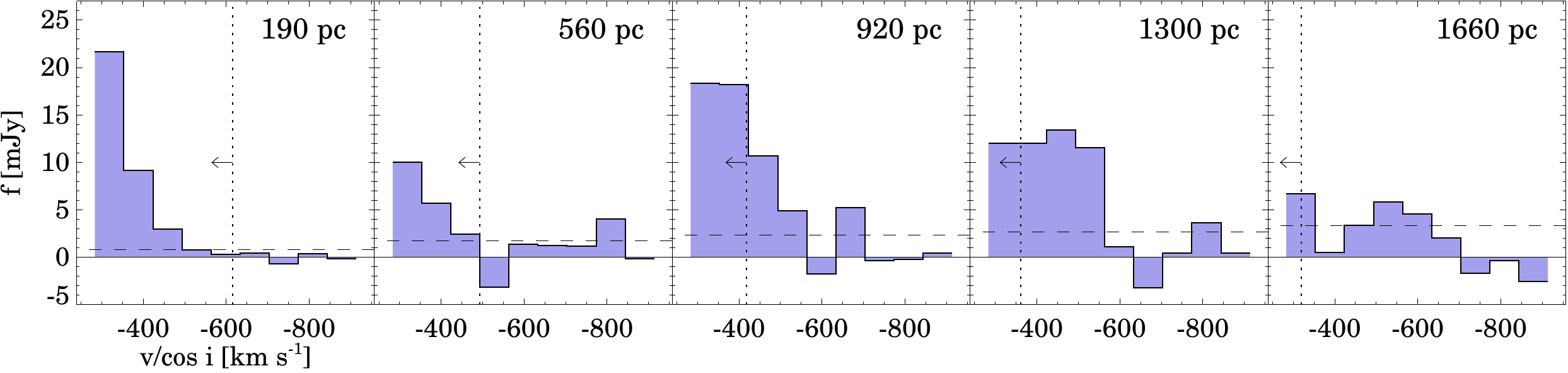}
\includegraphics[width=0.75\textwidth]{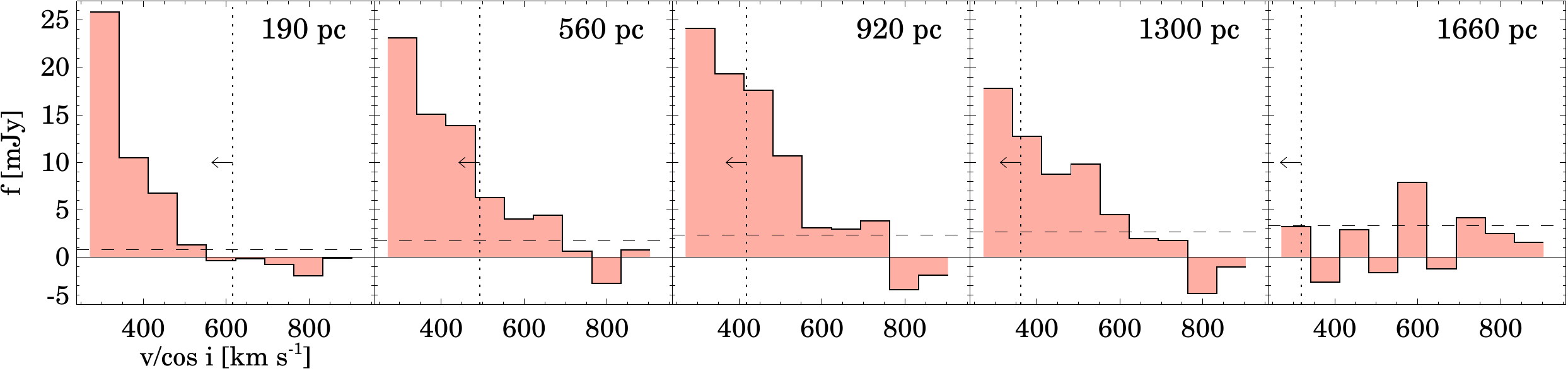}
\caption{High-velocity, $|$v$_{\rm proj}|>200$\,km\,s$^{-1}$, CO(2--1) outflow emission measured in the annulus sectors presented in Figures~\ref{fig:annuli_blue} and \ref{fig:annuli_red}. The top and bottom panels correspond to the blue- and red-shifted emissions of the outflow, respectively. The x-axis velocities and the distance to the nucleus indicated in each panel are deprojected values assuming a perpendicular to the disk outflow and a galaxy inclination, $i$, of 43\degree \citep{Pereira2016b}. The dotted vertical line traces the expected velocity of a cloud in the gravitational potential at each distance with an initial v\,$\sim600$\,km\,s$^{-1}$ at 190\,pc.
The horizontal dashed line indicates the average 1$\sigma$ noise of the spectra per channel.
\label{fig:sp_annulus_outflow}}
\end{figure*}

We use the spatially resolved information on the kinematics of the outflow to investigate the dynamical evolution of the high-velocity molecular gas. We extracted the total CO(2--1) emission in each of the annulus sectors defined in Section~\ref{ss:outflow_sled} (see also Figures~\ref{fig:annuli_blue} and \ref{fig:annuli_red}).
We chose the CO(2--1) transition since it has the higher signal-to-noise ratio. The high-velocity gas emission (projected $|$v$|>200$\,km\,s$^{-1}$) extracted in the B380 to B1130 and R380 to R1130 annulus sectors is shown in Figure~\ref{fig:sp_annulus_outflow}. We exclude the inner sectors since the broad nuclear emission prevents a clear separation between the outflow and the disk emissions. In this section, we use deprojected values for the velocities and distances. We assume a perpendicular to the disk outflow and a galaxy inclination of 43\degree\ (see \citealt{Pereira2016b}). 

\begin{figure}
\centering
\includegraphics[width=0.3\textwidth]{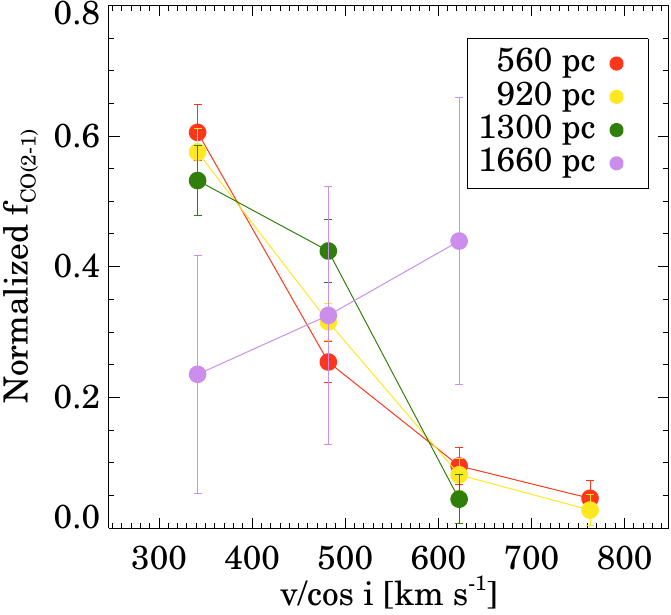}
\caption{Fraction of the high-velocity CO(2--1) emission as a function of the deprojected velocity for annulus sectors at various deprojected radii (see also Figure~\ref{fig:sp_annulus_outflow}).  We joined the data points to guide the eye.
\label{fig:outflow_norm_flux}}
\end{figure}

\subsubsection{Gravitational potential model}\label{ss:grav_pot}

To determine the dynamical evolution of the gas at sub-kpc spatial scale, we created a model of the gravitational potential. We modeled the stellar mass distribution in the central 14\arcsec (3.4\,kpc) using the near-IR F110W (J-band) and F160W (H-band) \textit{HST}\slash NICMOS images of this galaxy \citep{AAH06s}. First, we deprojected the images using a 43\degree\ inclination. Then, we created a light profile using 10 concentric annular apertures centered at the nucleus with radius steps of 0\farcs7 (180\,pc). The integrated fluxes are 46\,mJy and 75\,mJy in the F110W and F160W images, respectively. Finally, from the $J-H$ color, we estimated a mass-to-light ratio of 0.75 in solar units for the F160W luminosity from \citet{Bell2001}. Applying this ratio, we find a total stellar mass in the central 3.4\,kpc of 10$^{10.4}$\,\Msun\ which we assume is distributed in uniform annuli following the F160W light profile. The molecular gas in this area is 10$^{9.6}$\,\Msun, which is only 15\% of the stellar mass, and its contribution is not considered for the potential.

To account for the stellar mass at larger radii, we used the \textit{Spitzer}\slash IRAC 3.6\micron\ image. Similarly, we deprojected the 3.6\micron\ image and measured a flux of 85\,mJy on a 28\arcsec\ (6.7\,kpc) radius aperture. This corresponds to a stellar mas of 10$^{10.8}$\,\Msun\ using a mass-to-light ratio of 0.47 \citep{McGaugh2014}. For the mass distribution model, we assume that the extra stellar mass we measure in the IRAC 3.6\micron\ image with respect to the NICMOS images, $\sim$10$^{10.6}$\,\Msun\, is distributed in a uniform annulus with inner radius of 1.7\,kpc and outer radius of 6.7\,kpc.

This mass distribution reproduces relatively well the observed H$\alpha$ rotation velocity of v$_{\rm deproj}\sim$200\,km\,s$^{-1}$ at $\sim$5\,kpc (20\arcsec) away from the nucleus (see \citealt{Bellocchi2013}). Therefore, we do not include a dark matter halo potential since it is not expected to have an important effect on the outflow dynamics for the range of radii considered here ($<$2\,kpc). The halo affects the escape velocity and, therefore, the final fate of the outflowing gas (i.e, escape to the intergalactic medium or return to the galaxy disk).

The final potential is the sum of potentials produced by all these mass distributions and the gravitational acceleration its derivate along the $z$-axis. We only calculate the potential and acceleration along the $z$-axis perpendicular to the disk where we model the outflow movement.
The equations used are presented in Appendix~\ref{apx:gravity}.

\subsubsection{Dynamics of the outflowing molecular gas}

We used the gravitational potential derived before to investigate if the observed velocities are compatible with a pure gravitational evolution of the gas within this potential. While we cannot reject the pure gravity evolution, some aspects of the observations are not straightforward explained by this model.

In particular, the vertical dotted lines in Figure~\ref{fig:sp_annulus_outflow} trace the expected decreasing velocity of a cloud with an initial velocity of $\sim$600\,km\,s$^{-1}$ at 190\,pc (upper limit of the velocities detected at that distance). In a pure gravitational evolution scenario, the gas observed to the right of the dotted line must have had an initial velocity higher than $\sim$600\,km\,s$^{-1}$. However, we do not detect this high-velocity gas close to the nucleus.
For instance, the 700\,km\,s$^{-1}$ observed at a distance of 1660\,pc (560\,pc) was launched 1.6\,Myr (0.6\,Myr) ago with an initial velocity of 925\,km\,s$^{-1}$ (850\,km\,s$^{-1}$). So a noticeable amount of high-velocity gas (v$>$800\,km\,s$^{-1}$) should have been present at 190\,pc between 0.6 and 1.6\,Myr ago but now it is completely absent.
Also, the fraction of outflowing gas as a function of the velocity at different distances is approximately constant up to 1660\,pc (Figure~\ref{fig:outflow_norm_flux}). So to match the observed spatially resolved outflow velocity and mass structures, a combination of short timescale ($\sim$0.5\,Myr) varying mass outflow rates and initial outflow velocity distributions is needed. 
This outflow is launched by the obscured nuclear starburst (15\,\Msun\,yr$^{-1}$ and supernovae rate 0.2\,yr$^{-1}$) with abundant molecular gas available (depletion time $\sim$80\,Myr; \citealt{Pereira2016b}), so we do not expect the outflow launching processes (e.g., radiation from young stars, supernovae) to change drastically in $\sim$0.5\,Myr timescales.

Alternatively, these observations are easier to explain if the outflowing molecular gas is being accelerated. We distinguish two phases in this acceleration. First, the rotating gas in the nuclear region is accelerated until it is decoupled and leaves the plane of the galaxy. To get a rough estimate of this launching acceleration, we use the difference in the maximum outflow velocity at 190\,pc ($\sim$500\,km\,s$^{-1}$) and at 560\,pc ($\sim$800\,km\,s$^{-1}$). This yields a velocity gradient of $\sim$0.8\,km\,s$^{-1}$\,pc$^{-1}$.
This is comparable to the velocity gradient of the outflowing molecular gas observed in the outflow of NGC~253 (1\,km\,s$^{-1}$\,pc$^{-1}$), although this value was measured on smaller spatial scales of 100\,pc \hbox{\citep{Walter2017}.}
In M82, a velocity gradient of 32\,km\,s$^{-1}$\,pc$^{-1}$ is observed for the outflowing gas. But instead of the gas being accelerated, this gradient is interpreted as the blending of the rotation close to the disk and the double peaked profile expected for a conical outflow morphology \citep{Leroy2015}. In ESO~320-G030, a rotational component is not observed in the outflow region close to the disk, the outflow emission is not broader closer to the disk as in M82, and a double peaked profile is not seen (Figure~\ref{fig:sp_annulus_outflow}). So the geometry and dynamics of the outflow might vary depending on the object.

A second acceleration phase of the outflow would be needed to compensate the effect of gravitational attraction and explain why the maximum outflow velocity does not decrease and instead remains approximately constant up to 1.7\,kpc (Figure~\ref{fig:sp_annulus_outflow}). This larger scale acceleration is discussed in the next Section.

\subsubsection{Ram pressure acceleration and cloud destruction}\label{ss:ram_accel}

Two main mechanisms, radiation pressure and ram pressure, have been explored by models and simulations to explain the acceleration of the gas in outflows (e.g., \citealt{Murray2005}). For ESO~320-G030, at $>250$\,pc away from the nucleus, the effect of the radiation pressure is negligible compared with the gravitational attraction based on the gravitation potential derived in Section\,\ref{ss:grav_pot} and the IR luminosity of the galaxy ($g_{\rm grav}>20 g_{\rm rad}$ if $L_{\rm IR} = 10^{11.3}L_{\odot}$ and the average gas cross section, $\langle\sigma\rangle<10$\,cm$^2$\,g$^{-1}$ for $T_{\rm rad}\sim$100\,K; \citealt{Draine2011Rad}), so we focus on the ram pressure. The acceleration produced by ram pressure, $g_{\rm ram}$, is (e.g., \citealt{Schneider2017, Zhang2017}):
\begin{equation}
g_{\rm ram} = n_{\rm wind} ({\rm v}_{\rm wind} - {\rm v}_{\rm cold})^2 \frac{1}{N_{\rm H}}
\end{equation}

where v$_{\rm wind}$ and $n_{\rm wind}$ are the velocity and density of the hot wind material which depend on the distance $r$ to the nucleus, and v$_{\rm cold}$ and $N_{\rm H}$ the velocity and column density of the cold molecular clouds.

A molecular cloud in the outflow with a column density, $N_{\rm H}$, located at a distance $r$ from the nucleus will be affected by two opposing accelerations: ram pressure, $g_{\rm ram}(r)$, and gravity $g_{\rm grav}(r)$. If the density, $n_{\rm wind}$, and velocity, v$_{\rm wind}$, of the hot medium are high enough, there is an equilibrium velocity, v$_{\rm cold}$, of the cloud at which both accelerations cancel out. The equilibrium velocity decreases with increasing $N_{\rm H}$. Consequently, the observed velocity distribution of the molecular outflowing gas at a given radius would trace the $N_{\rm H}$ distribution of the cold clouds at that position.

Cold clouds evaporate in hot winds, reducing their column density, and eventually completely dissipate (e.g., \citealt{Scannapieco2015, Armillotta2017, Zhang2017, Sparre2019}). Simulations also show that clouds with lower $N_{\rm H}$ are easier to destroy \citep{Bruggen2016}. However, as we discussed in Section~\ref{ss:thermal_ev}, the heating of the outflowing molecular gas is not efficient and this facilitates the survival of the cold clouds. In the case of ESO~320-G030 molecular gas is detected up to $\sim$1.6\,kpc. This radius is similar to the molecular outflow radii from hundreds of pc to $\sim$1\,kpc measured in other star-forming galaxies (e.g., \citealt{Fluetsch2019}). Therefore, the survival of cold molecular gas embedded in hot winds seems to be limited to relatively brief timescales of few Myr given the typical velocities of the molecular phase of 100--800\,km\,s$^{-1}$. In \citet{Pereira2016b}, we analyzed the molecular clumps in the outflow of ESO~320-G030. Their mass decreases at larger distances and this would be consistent with the evaporation of the clumps within the hot wind.

Based on these two properties (that the observed velocity depends on $N_{\rm H}$ and that the dissipation of the cold clouds within hot winds is faster at low $N_{\rm H}$), we can explain the observations of the molecular phase of the ESO~320-G030 outflow:
(1) The constant maximum velocity of $\sim$700--800\,km\,s$^{-1}$ between 560 and 1660\,pc would be given by the lowest $N_{\rm H}$ that can efficiently survive in the outflow at each distance; and (2) clouds would be losing mass and decreasing their $N_{\rm H}$ as they travel away from the galaxy disk. So, the maximum distance at which molecular gas is detected would depend on the highest $N_{\rm H}$ that can be launched at the base of the outflow.
As a consequence of all this, an approximately constant outflow rate during timescales of tens to hundred Myr, as expected from an intense starburst, would present similar short flow times (i.e., $R_{\rm max}\slash v_{\rm cold}$) of few Myr.

After the destruction of the molecular clouds, it is likely that the gas becomes atomic neutral. In ESO~320-G030, the atomic neutral phase of the outflow has been detected through the optical NaD absorption \citep{Cazzoli2014, Cazzoli2016}. In addition, there is evidence of a warm ($T\sim10^4$\,K) ionized phase of the outflow traced by a broad H$\alpha$ profile \citep{Arribas2014}. 
Understanding the relation between the different outflow phases will help to further investigate the evolution of the outflowing molecular gas. To do so, optical\slash IR integral field spectroscopy observations of the atomic ionized and neutral phases at angular resolutions comparable to that of ALMA would be necessary.

\section{Conclusions}\label{s:conclusions}

We studied the spatially resolved excitation and acceleration of the cold molecular outflow of the nearby starburst LIRG ESO~320-G030 ($d$=48\,Mpc, $L_{\rm IR}\slash L_\odot$=10$^{11.3}$) using ALMA 70\,pc resolution CO multi-transition (1--0, 2--1, 4--3, and 6--5) data. ESO~320-G030 is a double-barred isolated spiral, but its compact (250\,pc) and extremely obscured nuclear starburst (SFR$\sim$15\,\Msun\,yr$^{-1}$; $A_{\rm V}$\,$\sim$40\,mag) as well as its very high nuclear molecular gas surface density ($\Sigma_{\rm H_2} = 10^{4.4}$\,$M_\odot$\,pc$^{-2}$) resemble those of more luminous starburst-dominated ULIRGs.
We determined the physical conditions of the outflowing molecular gas at different distances from the launching site by fitting the CO SLEDs with non-LTE radiative transfer models. We also studied the spatially resolved kinematics of the high-velocity molecular gas and modeled its dynamical evolution under the effects of the gravitational potential and ram pressure due to the hot wind environment. The main results of this work are the following:
\begin{enumerate}
\item The outflowing molecular gas is less excited than the molecular gas in the nucleus (i.e., outflow launching site) and the disk GMCs. The \hbox{CO(1--0)\slash CO(2--1)} ratio is enhanced in the outflow with respect to the disk and nucleus, as already observed in other starburst outflows, and the CO(4--3) transition is not detected in the outflow beyond 250\,pc with a robust upper limit. This lower outflow excitation with respect to the disk is also opposite to what has been observed in most AGN-driven molecular outflows in the literature.

\item The non-LTE radiative transfer modeling of the CO SLEDs shows that the molecular clouds in the outflow have lower kinetic temperature ($\sim$9\,K), column densities ($N_{\rm CO}\slash \Delta$v), filling factors (e.g., $\Phi \Delta$v), and smaller sizes (or lower CO abundances) than the molecular clouds in the disk and the nucleus. The high velocity gradient in the outflowing clouds, $(\Delta$v$\slash R)_{\rm LVG}$, results in $K_{\rm vir}$ in the range {2--6000}. This indicates that they are not gravitationally bound, unless the CO abundance is {at least$>2$} times lower in the outflow than the assumed $Z_{\rm CO}$ = 10$^{-4}$. Thus, the different physical conditions suggest that the heating\slash cooling, stability, collapse and chemical composition of the outflowing clouds differ from those of the clouds in the disk of the galaxy and, therefore, outflowing clouds might not be able to form stars.

\item {We estimate that the hotter molecular gas in the nucleus ($T_{\rm kin}=$38$^{+119}_{-21}$) can cool down to the outflow molecular gas temperature, 9\,K, in $\sim$0.1\,Myr. Therefore, these observations are compatible with the outflowing molecular gas being molecular gas from the galaxy entrained and accelerated. However, we cannot reject that the outflowing molecular gas condensates in the outflow based on these data.}

\item The outflowing molecular gas low kinetic temperature, about 9\,K, is approximately constant between radial distances of 300 and 1700\,pc. This implies that the heating of the molecular gas by the hot wind environment is not efficient and this may help to increase the survival of the molecular clouds in the outflow hot medium. 

\item The spatially resolved kinematics of the molecular gas in the outflow shows that the maximum velocity as well as the velocity distribution are relatively uniform at radii between 600 and 1700\,pc. Also a $\sim$0.8\,km\,s$^{-1}$\,pc$^{-1}$ velocity gradient is estimated between 190\,pc and 560\,pc. The observed velocities could be explained just by pure gravitational evolution of the outflow, but would require a changing combination of mass outflow rate and initial velocity distribution of the gas. 

\item Alternatively, a favored scenario is a combination of ram pressure and inefficient cloud evaporation by the surrounding hot outflowing gas which could explain the observed kinematics and extent of the molecular phase of the outflow. The velocity distribution would be determined by: (i) the equilibrium velocity between ram pressure and gravity; and (ii) the $N_{\rm H}$ distribution of the outflowing clouds, with increasing velocities for decreasing $N_{\rm H}$. The maximum velocity would be given by the minimum $N_{\rm H}$ that can survive in the hot wind. The outflow size would be determined by the survival time of the molecular clouds with the highest $N_{\rm H}$ at the launching site. 
\end{enumerate}

\begin{acknowledgements}
{ We thank the referee for their comments and suggestions.}
We thank M. Villar-Mart\'in for useful comments and careful reading of the manuscript.
MPS acknowledges support from the Comunidad de Madrid through the Atracci\'on de Talento Investigador Grant 2018-T1/TIC-11035, PID2019-105423GA-I00 (MCIU/AEI/FEDER,UE), and STFC through grants ST/N000919/1 and ST/N002717/1. LC acknowledge financial support by the Spanish MICINN under grant ESP2017-83197.
AA-H, SG-B, and AU acknowledge support through grant PGC2018-094671-B-I00 (MCIU/AEI/FEDER,UE). 
E.GA is a Research Associate at the Harvard-Smithsonian Center for Astrophysics, and thanks the Spanish Ministerio de Econom\'{\i}a y Competitividad for support under project ESP2017-86582-C4-1-R.
JP-L acknowledge financial support by the Spanish MICINN under grant AYA2017-85170-R.
MPS, AA-H, LC, JP-L and SA work was done under project No. MDM-2017-0737 Unidad de Excelencia "Mar\'ia de Maeztu"- Centro de Astrobiología (INTA-CSIC).
DR acknowledges support from ST/S000488/1 and the University of Oxford Fell Fund.
This paper makes use of the following ALMA data: ADS/JAO.ALMA\#2016.1.00263.S, ADS/JAO.ALMA\#2013.1.00271.S. ALMA is a partnership of ESO (representing its member states), NSF (USA) and NINS (Japan), together with NRC (Canada) and NSC and ASIAA (Taiwan) and KASI (Republic of Korea), in cooperation with the Republic of Chile. The Joint ALMA Observatory is operated by ESO, AUI/NRAO and NAOJ.
The National Radio Astronomy Observatory is a facility of the National Science Foundation operated under cooperative agreement by Associated Universities, Inc.

\end{acknowledgements}
\bibliographystyle{aa}

\appendix
\section{Gravitational potential and acceleration}\label{apx:gravity}
The potential produced by a thin disk along the $z$-axis (i.e., $r=0$) where the outflow evolution is modeled is the following (e.g., \citealt{Lass1983}):
\begin{equation}
 V(z) = 2\pi G \sigma \left[|z| - \sqrt{R^2 + z^2}\right],
\end{equation}
where $G$ is the gravitational constant, $\sigma = M\slash (\pi R^2)$ is the mass surface density of the disk, and $R$ the radius of the disk.
For a thin annular disk, the potential is the difference between potential of the larger disk and that of the smaller one:
\begin{equation}
 V(z) = \frac{G M}{a w} \left[ \sqrt{(a - w\slash 2)^2 + z^2} - \sqrt{(a + w\slash 2)^2 + z^2} \right],
\end{equation}
where $M$ is the mass of the annular disk, $w = a_{\rm out} - a_{\rm in}$ the width of the ring, $a= 0.5\times(a_{\rm out} + a_{\rm in})$ the average radius, and $a_{\rm in}$ and $a_{\rm out}$ are the inner and outer radii of the annular disk, respectively.

The acceleration produced by this potentials along the $z$-axis is $-{\rm d}V\slash {\rm d}z$. For the disk, this corresponds to:
\begin{equation}
 g(z) = -2\pi G \sigma \left[1 - \frac{z}{\sqrt{R^2 + z^2}}\right],
\end{equation}
and for the annular disk:

\begin{equation}
 g(z) = -\frac{G M}{a w} \left[ \frac{z}{\sqrt{(a - w\slash 2)^2 + z^2}} - \frac{z}{\sqrt{(a + w\slash 2)^2 + z^2}} \right].
\end{equation}

\end{document}